\documentclass[12pt,a4paper]{article}

\usepackage{mathtools}
\usepackage{epsfig}
\usepackage{lineno}  
\usepackage{graphicx}  
\usepackage{xspace} 
\usepackage{color}
\usepackage{colortbl}
\usepackage{amsmath} 
\usepackage{ifthen} 

\newboolean{pdflatex}
\setboolean{pdflatex}{false} 
\newboolean{articletitles}
\setboolean{articletitles}{true} 

\newboolean{uprightparticles}
\setboolean{uprightparticles}{false} 

\usepackage{amssymb}
\usepackage{amsfonts}
\usepackage{upgreek} 

\usepackage{hyperref}    
\usepackage[all]{hypcap} 

\def\lhcb {LHCb\xspace}
\def\ux85 {UX85\xspace}

\ifthenelse{\boolean{uprightparticles}}%
{

 \def\Ppsi        {\ensuremath{\uppsi}\xspace}

 \def\PDelta      {\ensuremath{\Delta}\xspace}                 
 \def\PXi      {\ensuremath{\Xi}\xspace}                 
 \def\PLambda      {\ensuremath{\Lambda}\xspace}                 
 \def\PSigma      {\ensuremath{\Sigma}\xspace}                 
 \def\POmega      {\ensuremath{\Omega}\xspace}                 
 \def\PUpsilon      {\ensuremath{\Upsilon}\xspace}                 
 
 \def\PB      {\ensuremath{\mathrm{B}}\xspace}                 
                  
 \def\PD      {\ensuremath{\mathrm{D}}\xspace}

 \def\PJ      {\ensuremath{\mathrm{J}}\xspace}                 
 \def\PK      {\ensuremath{\mathrm{K}}\xspace}

 \def\Pb      {\ensuremath{\mathrm{b}}\xspace}                 
 \def\Pc      {\ensuremath{\mathrm{c}}\xspace}

 \def\Pi      {\ensuremath{\mathrm{i}}\xspace}

}
{

 \def\Ppsi        {\ensuremath{\psi}\xspace}                 
                  
 \mathchardef\PDelta="7101
 \mathchardef\PXi="7104
 \mathchardef\PLambda="7103
 \mathchardef\PSigma="7106
 \mathchardef\POmega="710A
 \mathchardef\PUpsilon="7107
                  
 \def\PB      {\ensuremath{B}\xspace}                 
                  
 \def\PD      {\ensuremath{D}\xspace}

 \def\PJ      {\ensuremath{J}\xspace}                 
 \def\PK      {\ensuremath{K}\xspace}

 \def\Pb      {\ensuremath{b}\xspace}                 
 \def\Pc      {\ensuremath{c}\xspace}

 \def\Pi      {\ensuremath{i}\xspace}

}

\def\cquark    {\ensuremath{\Pc}\xspace}

\def\bquark    {\ensuremath{\Pb}\xspace}

\def\kaon  {\ensuremath{\PK}\xspace}
  \def\Kbar  {\kern 0.2em\overline{\kern -0.2em \PK}{}\xspace}

\def\Kz    {\ensuremath{\kaon^0}\xspace}
\def\Kzb   {\ensuremath{\Kbar^0}\xspace}
\def\KzKzb {\ensuremath{\Kz \kern -0.16em \Kzb}\xspace}
\def\Kp    {\ensuremath{\kaon^+}\xspace}
\def\Km    {\ensuremath{\kaon^-}\xspace}

\def\KpKm  {\ensuremath{\Kp \kern -0.16em \Km}\xspace}

  \def\Dbar    {\kern 0.2em\overline{\kern -0.2em \PD}{}\xspace}
\def\D       {\ensuremath{\PD}\xspace}

\def\Dz      {\ensuremath{\D^0}\xspace}
\def\Dzb     {\ensuremath{\Dbar^0}\xspace}
\def\DzDzb   {\ensuremath{\Dz {\kern -0.16em \Dzb}}\xspace}
\def\Dp      {\ensuremath{\D^+}\xspace}
\def\Dm      {\ensuremath{\D^-}\xspace}

\def\DpDm    {\ensuremath{\Dp {\kern -0.16em \Dm}}\xspace}

  \def\Bbar    {\kern 0.18em\overline{\kern -0.18em \PB}{}\xspace}

\def\jpsi     {\ensuremath{{\PJ\mskip -3mu/\mskip -2mu\Ppsi\mskip 2mu}}\xspace}

  \def\Y#1S{\ensuremath{\PUpsilon{(#1S)}}\xspace}

\def\ra                 {\ensuremath{\rightarrow}\xspace}
\def\to                 {\ensuremath{\rightarrow}\xspace}

\def\AT#1     {\ensuremath{A_{\mathrm{T}}^{#1}}\xspace}           

\def\C#1      {\ensuremath{\mathcal{C}_{#1}}\xspace}                       
\def\Cp#1     {\ensuremath{\mathcal{C}_{#1}^{'}}\xspace}                    
\def\Ceff#1   {\ensuremath{\mathcal{C}_{#1}^{\mathrm{(eff)}}}\xspace}        
\def\Cpeff#1  {\ensuremath{\mathcal{C}_{#1}^{'\mathrm{(eff)}}}\xspace}       
\def\Ope#1    {\ensuremath{\mathcal{O}_{#1}}\xspace}                       
\def\Opep#1   {\ensuremath{\mathcal{O}_{#1}^{'}}\xspace}                    

\newcommand{\tev}{\ensuremath{\mathrm{\,Te\kern -0.1em V}}\xspace}
\newcommand{\gev}{\ensuremath{\mathrm{\,Ge\kern -0.1em V}}\xspace}
\newcommand{\mev}{\ensuremath{\mathrm{\,Me\kern -0.1em V}}\xspace}
\newcommand{\kev}{\ensuremath{\mathrm{\,ke\kern -0.1em V}}\xspace}
\newcommand{\ev}{\ensuremath{\mathrm{\,e\kern -0.1em V}}\xspace}
\newcommand{\gevc}{\ensuremath{{\mathrm{\,Ge\kern -0.1em V\!/}c}}\xspace}
\newcommand{\mevc}{\ensuremath{{\mathrm{\,Me\kern -0.1em V\!/}c}}\xspace}
\newcommand{\gevcc}{\ensuremath{{\mathrm{\,Ge\kern -0.1em V\!/}c^2}}\xspace}
\newcommand{\gevgevcccc}{\ensuremath{{\mathrm{\,Ge\kern -0.1em V^2\!/}c^4}}\xspace}
\newcommand{\mevcc}{\ensuremath{{\mathrm{\,Me\kern -0.1em V\!/}c^2}}\xspace}

\def\mum  {\ensuremath{\,\upmu\rm m}\xspace}

\def\invpb {\ensuremath{\mbox{\,pb}^{-1}}\xspace}

\def\gsim{{~\raise.15em\hbox{$>$}\kern-.85em
          \lower.35em\hbox{$\sim$}~}\xspace}
\def\lsim{{~\raise.15em\hbox{$<$}\kern-.85em
          \lower.35em\hbox{$\sim$}~}\xspace}

\def\pt         {\mbox{$p_{\rm T}$}\xspace}

\newcommand{\lum} {\ensuremath{\mathcal{L}}\xspace}

\def\tell1  {TELL1\xspace}
\def\ukl1   {UKL1\xspace}

\newcommand{\bit}{\begin{itemize}}
\newcommand{\eit}{\end{itemize}}
\newcommand{\bce}{\begin{center}}
\newcommand{\beqn}{\begin{eqnarray*}}
\newcommand{\eeqn}{\end{eqnarray*}}
\newcommand{\beq}{\begin{equation}}
\newcommand{\ece}{\end{center}}
\newcommand{\upsmm}{\nobreak{\ensuremath{\varUpsilon\rightarrow \mu^+\mu^-}}}
\newcommand{\ups}{\ensuremath{\varUpsilon}}
\newcommand{\ones}{\ensuremath{\varUpsilon(1S)}}
\newcommand{\twos}{\ensuremath{\varUpsilon(2S)}}
\newcommand{\threes}{\ensuremath{\varUpsilon(3S)}}
\newcommand{\ismm}{\ensuremath{\varUpsilon(iS)\rightarrow\mu^{+}\mu^{-}}}
\newcommand{\onesmm}{\ensuremath{\varUpsilon(1S)\rightarrow\mu^+\mu^-}}
\newcommand{\twosmm}{\ensuremath{\varUpsilon(2S)\rightarrow\mu^+\mu^-}}
\newcommand{\threesmm}{\ensuremath{\varUpsilon(3S)\rightarrow\mu^+\mu^-}}

\newcommand{\bmmis}{\ensuremath{\mathcal{B}(\varUpsilon(iS)\rightarrow\mu^+\mu^-)}}
\newcommand{\bmmones}{\ensuremath{\mathcal{B}(\varUpsilon(1S)\rightarrow\mu^+\mu^-)}}
\newcommand{\bmmtwos}{\ensuremath{\mathcal{B}(\varUpsilon(2S)\rightarrow\mu^+\mu^-)}}
\newcommand{\bmmthrees}{\ensuremath{\mathcal{B}(\varUpsilon(3S)\rightarrow\mu^+\mu^-)}}

\usepackage{cite}
\usepackage{mciteplus}
\begin{document}



\begin{titlepage}
\pagenumbering{roman}

\vspace*{-1.5cm}
\centerline{\large EUROPEAN ORGANIZATION FOR NUCLEAR RESEARCH (CERN)}
\vspace*{1.5cm}
\hspace*{-0.5cm}
\begin{tabular*}{\linewidth}{lc@{\extracolsep{\fill}}r}
\ifthenelse{\boolean{pdflatex}}
{\vspace*{-2.7cm}\mbox{\!\!\!\includegraphics[width=.14\textwidth]{figs/lhcb-logo.pdf}} & &}%
{\vspace*{-1.2cm}\mbox{\!\!\!\includegraphics[width=.12\textwidth]{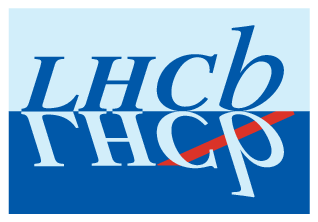}} & &}%
\\
 & & CERN-PH-EP-2012-051 \\  
 & & LHCb-PAPER-2011-036 \\  
\end{tabular*}

\vspace*{4.0cm}

{\bf\boldmath\huge
\begin{center} 
\boldmath Measurement of $\ups$ production
in $pp$ collisions at $\sqrt{s}=7~\tev$
\end{center}
}

\centering{
The LHCb collaboration
\footnote{Authors are listed on the following pages.}}

\begin{abstract}
  \noindent
The production of $\ones$, $\twos$ and $\threes$ mesons in proton-proton collisions at the 
centre-of-mass energy of ${\sqrt{s}=7~\tev}$ is studied with the LHCb detector. 
The analysis is based on a data sample of $25~\invpb$ 
collected at the Large Hadron Collider.
The $\ups$ mesons are reconstructed in the decay mode $\upsmm$
and the signal yields are extracted from a fit to the $\mu^+\mu^-$ invariant 
mass distributions.
The differential production cross-sections times dimuon branching fractions are measured 
as a function of the $\ups$ 
transverse momentum $\pt$ and rapidity $y$,  over the range 
$\pt<15~\gevc$ and $2.0<y<4.5$. 
The cross-sections times branching fractions, integrated 
over these kinematic ranges, are measured to be 
%
\begin{alignat*}{1}
\sigma(pp\ra \ones\,   X)\times\bmmones = 2.29\phantom{0}  \pm  0.01\phantom{0}   \pm 0.10\phantom{0}\,\, _{-0.37}^{+0.19}~{\rm nb}, \\
\sigma(pp\ra \twos\,   X)\times\bmmtwos = 0.562  \pm  0.007   \pm 0.023\, _{-0.092}^{+0.048}~{\rm nb}, \\
\sigma(pp\ra \threes\, X)\times\bmmthrees = 0.283  \pm  0.005 \pm 0.012 \,_{-0.048}^{+0.025}~{\rm nb},
\end{alignat*}
where the first uncertainty is 
statistical, the second systematic and the third is due to the unknown polarisation of the 
three $\ups$ states.
\end{abstract}
\vspace{1mm}
\centering{Published in Eur.~Phys.~J.~C volume 72,6 (June 2012)}
\end{titlepage}


\newpage
\setcounter{page}{2}
\mbox{~}

\begin{flushleft}
\begin{center}{{\bf \lhcb collaboration }}\end{center}
\vspace{0.2cm}
R.~Aaij$^{38}$, 
C.~Abellan~Beteta$^{33,n}$, 
B.~Adeva$^{34}$, 
M.~Adinolfi$^{43}$, 
C.~Adrover$^{6}$, 
A.~Affolder$^{49}$, 
Z.~Ajaltouni$^{5}$, 
J.~Albrecht$^{35}$, 
F.~Alessio$^{35}$, 
M.~Alexander$^{48}$
% ...

{\footnotesize \it
% ...
$ ^{5}$Clermont Universit\'{e}, Universit\'{e} Blaise Pascal, CNRS/IN2P3, LPC, Clermont-Ferrand, France\\
$ ^{6}$CPPM, Aix-Marseille Universit\'{e}, CNRS/IN2P3, Marseille, France\\
% ...
$ ^{33}$Universitat de Barcelona, Barcelona, Spain\\
$ ^{34}$Universidad de Santiago de Compostela, Santiago de Compostela, Spain\\
$ ^{35}$European Organization for Nuclear Research (CERN), Geneva, Switzerland\\
% ...
$ ^{38}$Nikhef National Institute for Subatomic Physics, Amsterdam, The Netherlands\\
% ...
$ ^{43}$H.H. Wills Physics Laboratory, University of Bristol, Bristol, United Kingdom\\
% ...
$ ^{48}$School of Physics and Astronomy, University of Glasgow, Glasgow, United Kingdom\\
$ ^{49}$Oliver Lodge Laboratory, University of Liverpool, Liverpool, United Kingdom\\
% ...
$ ^{n}$LIFAELS, La Salle, Universitat Ramon Llull, Barcelona, Spain\\
% ...
}
\bigskip
\end{flushleft}





\pagestyle{plain} 
\setcounter{page}{1}
\pagenumbering{arabic}


\section{Introduction}\label{Introduction}

The measurement of heavy quark production in hadron collisions probes the dynamics of the colliding partons.
The study of heavy quark-antiquark resonances, such as the
 $b\overline{b}$ bound states $\ones$,
$\twos$ and $\threes$  (indicated generically as $\ups$ in the following)
is of interest as these mesons have large production 
cross-sections 
and can be produced in different spin configurations. 
In addition, the thorough understanding of these states is the first step 
towards the study of recently discovered new states
in the $b\bar{b}$ system~\cite{exotic1,exotic2,exotic3,chib3p}.
Although \ups\ production was studied by 
several experiments 
in the past, the underlying production mecha\-nism is still not well understood.
Several models exist 
but fail to reproduce both the cross-section and the 
polarisation measurements at the Tevatron~\cite{cdf1, d01, d02}.
Among these  are the 
Colour Singlet Model (CSM)~\cite{CSM,CSM1,CSM2}, recently improved by adding 
higher order contributions (NLO CSM), 
the standard truncation of the nonrelativistic QCD expansion (NRQCD)~\cite{COM1}, 
which includes contributions from the Colour Octet Mechanism~\cite{COM2,COM3},
and the Colour Evaporation Model (CEM)~\cite{CEM}.
Although the disagreement of the theory with the data is less pronounced 
for  bottomonium than for charmonium, the measurement of $\ups$
production 
is  important as the theoretical calculations are 
more robust due to the heavier bottom quark.

There are two major sources of \ups\ production in $pp$ collisions: 
direct production 
and feed-down from the decay of heavier prompt bottomonium states, 
like $\chi_{b}$, 
or higher-mass \ups\ states.
This study presents  measurements of the individual inclusive production cross-sections 
of the three $\ups$ mesons decaying into a pair of muons.
The measurements are performed in $7~\tev$ 
centre-of-mass $pp$ collisions 
as a function of the $\ups$ 
transverse momentum ($\pt<15~\gevc$) and rapidity ($2<y<4.5$), 
in 15 bins of $\pt$ and five bins of $y$.
This analysis is 
complementary to those recently presented by the 
ATLAS
collaboration, who measured the $\ones$ cross section for
$|y|<2.4$~\cite{atlas}, and the CMS
collaboration, who measured the $\ones,\twos$ and $\threes$
cross sections in the rapidity 
region $|y|<2.0$~\cite{cms}.

\section{The LHCb detector and data}\label{detector}

The results presented here are based on a dataset of $25.0\pm 0.9~\invpb$
collected at the Large Hadron Collider (LHC) in 2010 with the LHCb detector
at a centre-of-mass energy of 7~\tev.

The \lhcb detector~\cite{lhcbdetectorpaper} is a single-arm forward
spectrometer covering the pseudo-rapidity range $2<\eta <5$, designed
for the study of particles containing \bquark or \cquark quarks. The
detector includes a high precision tracking system consisting of a
silicon-strip vertex detector surrounding the $pp$ interaction region, 
a large-area silicon-strip detector located upstream of a dipole
magnet with a bending power of about $4{\rm\,Tm}$, and three stations
of silicon-strip detectors and straw drift-tubes placed
downstream. The combined tracking system has a momentum resolution
$\Delta p/p$ that varies from 0.4\% at 5\gevc to 0.6\% at 100\gevc,
and an impact parameter resolution of 20\mum for tracks with high
transverse momentum. Charged hadrons are identified using two
ring-imaging Cherenkov detectors. Photon, electron and hadron
candidates are identified by a calorimeter system consisting of
scintillating-pad and pre-shower detectors, an electromagnetic
calorimeter and a hadronic calorimeter. Muons are identified by 
a muon
system composed of alternating layers of iron and multiwire
proportional chambers. 
The trigger consists of a hardware stage,
 based
on information from the calorimeter and muon systems, followed by a
software stage which applies a full event reconstruction.
This analysis uses events triggered by one or two muons. 
At the hardware level one or two muon candidates are required 
with $\pt$ larger than 1.4~\gevc for one muon, and 0.56 and 
0.48~\gevc
for two muons.
At the software level, 
the combined dimuon mass is required to be greater than 2.9~\gevcc,
and both the tracks and the vertex have to be of good quality.
To avoid the possibility that a few events with a high occupancy dominate the
trigger processing time, 
a set of global event selection requirements 
based on hit multiplicities is applied.

The Monte Carlo samples used are based on the {\sc Pythia} 6.4 
generator~\cite{pythia},
with a choice of parameters specifically configured for 
LHCb~\cite{procgenerator}. 
The {\sc EvtGen} package~\cite{EvtGen} describes the 
decay of the \ups\ resonances, 
and the {\sc Geant4} package~\cite{Geant4} simulates the detector 
response.
The prompt bottomonium production processes activated in {\sc Pythia} 
are those from the leading-order
colour-singlet and colour-octet mechanisms for the $\ones$, and 
colour-singlet only for the $\twos$ and the $\threes$.
QED radiative corrections to the decay $\upsmm$ are generated
with the {\sc Photos} 
package~\cite{photos}.

\section{Cross-section determination}

\begin{figure}[!bt]
\bce
 \includegraphics[width=0.6\textwidth]{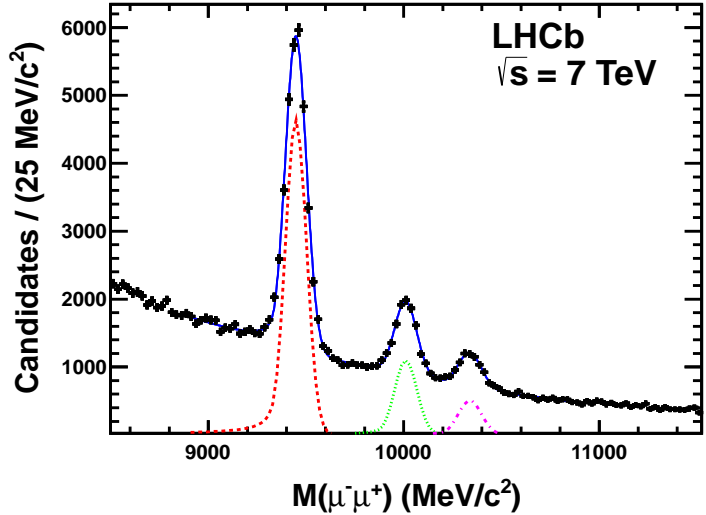}
 \caption{\small{Invariant mass distribution of the selected $\ups\to\mu^+\mu^-$ candidates
in the range $\pt<15~\gevc$ and $2.0<y<4.5$.  
The three peaks correspond to the $\ones$, $\twos$ and $\threes$ signals (from left to right). 
The superimposed curves are the result of the fit as described in the text.
}}\label{fig::massfit}
\ece
\end{figure}

The double differential cross-section for the inclusive $\ups$
 production of the three different states is computed as
\begin{equation}\label{eq::sigma}
\frac{{\rm d}^2\sigma^{iS}}{{\rm d}\pt{\rm d}y}\,\times\,
\mathcal{B}^{iS}= \frac{N^{iS}}
{\lum \times \varepsilon^{iS} \times\Delta y\times\Delta \pt},\quad i=1,2,3;
\end{equation}
where $\sigma^{iS}$ is the inclusive cross section $\sigma(pp\ra\ups(iS)X)$, $\mathcal{B}^{iS}$ is the 
dimuon branching fraction $\bmmis$, 
$N^{iS}$ is the number of observed $\ismm$ decays
in a given bin of $\pt$ and $y$, 
$\varepsilon^{iS}$ is the $\ismm$ total detection efficiency including acceptance effects, 
$\lum$ is the integrated luminosity 
and $\Delta y=0.5$ and $\Delta \pt=1~{\rm GeV}/c$ are the rapidity and $\pt$ bin sizes, respectively.
In order to estimate  $N^{iS}$, 
a fit to the reconstructed invariant mass distribution 
is performed  in each of the 15~$\pt\times 5~y$ bins.
$\ups$ candidates are formed from pairs of oppositely charged muon tracks 
which  traverse the full spectrometer
and satisfy the trigger requirements.
Each track must have $\pt>1~\gevc$, be identified as a muon and 
have a good quality of the track fit.  The two muons are required to originate from a common vertex
with a good $\chi^2$ probability. 
The three $\ups$ signal yields are determined from a fit to the reconstructed invariant 
mass $m$ of the selected $\ups$ candidates in the interval 
8.9--10.9~$\gevcc$. 
The mass distribution is described by a sum of three Crystal Ball functions~\cite{CB} for the $\ones$, $\twos$ and 
$\threes$ signals and an exponential function 
for the combinatorial background.
The Crystal Ball function is defined as
\begin{equation}
 f_{\mathrm{CB}} =
 \begin{dcases}
     \frac{\Big(\frac{n}{|a|}\Big)^n e^{-\frac{1}{2}a^2}}{\Big(\frac{n}{|a|}-|a|-\frac{m-M}{\sigma}\Big)^n} & {\mathrm{ if}}\,\,\, \frac{m-M}{\sigma} < -|a|\\
     \exp\Bigg( -\frac{1}{2}\Big(\frac{m-M}{\sigma}\Big)^2\Bigg) & {\mathrm {otherwise}},\label{eq::cball}
  \end{dcases}
\end{equation}
with  $f_{\mathrm{CB}} = f_{\mathrm{CB}}(m;M,\sigma,a,n)$,
where $M$ and $\sigma$ are the mean and width of the gaussian.
The parameters $a$ and $n$ 
describing the radiative tail of the $\ups$ 
mass distribution are fixed 
to describe a tail dominated by
QED photon emission, as confirmed by simulation.
The distribution in Fig.~\ref{fig::massfit} shows the results of the fit performed 
in the full range of $\pt$ and $y$. 
The signal yields 
obtained from the fit are $\ones=26\,410\pm 212$,
$\twos=6726 \pm 142$ and 
$\threes=3260\pm 112$ events.
The mass resolution of the 
 $\ones$ peak is $\sigma=53.9\pm0.5~\mevcc$.
The  resolutions of the $\twos$ and $\threes$ peaks are 
fixed to the resolution of the $\ones$,
scaled by the 
ratio of the masses, as expected from 
 resolution effects. The masses are 
allowed to vary in the fit and are measured to be
$M(\ones) = 9448.3\pm0.5$~\mevcc,
$M(\twos) = 10\,010.4\pm1.4$~\mevcc and
$M(\threes)=10\,338.7\pm2.6$~\mevcc, where 
the quoted uncertainties are statistical only.
The fit is repeated independently for each of the bins in
$p_{\rm T}$ and $y$. 
When fitting the individual bins, due to the 
reduced dataset, the masses and widths of the three $\ups$ states in the fit are fixed to 
the values obtained when fitting the full range.
Bins with fewer than 36 entries are excluded from the analysis.
The total efficiency $\varepsilon$ entering the cross-section expression of 
Eq.~\eqref{eq::sigma} is the product of 
the geometric acceptance, the reconstruction and selection 
efficiency and the trigger efficiency.
All efficiency terms have been evaluated using Monte Carlo 
simulations in each  ($\pt, y$) bin separately, with the 
exception of that related to the global event selection which has 
been determined from data.
In the simulation the $\ups$ meson is produced 
in an unpolarised state. 
The absolute luminosity scale was measured at specific periods 
during the 2010 data taking
using both van der Meer scans and a beam-gas imaging method~\cite{massi,jaap}.
The uncertainty on the integrated luminosity for the analysed 
sample due to this method is estimated~to~be~3.5\%~\cite{jaap}.
The knowledge of the absolute luminosity scale is used to calibrate the number
of tracks in the vertex detector, which is found to be stable throughout the
data-taking period and can therefore be used to monitor the instantaneous luminosity
of the entire data sample. The integrated luminosity of the data sample used in this
analysis is determined to be $25.0~\invpb$.

\section{Systematic uncertainties}

\begin{figure}[htb!]
\bce
 \includegraphics[width=0.45\textwidth, height=0.45\textwidth]{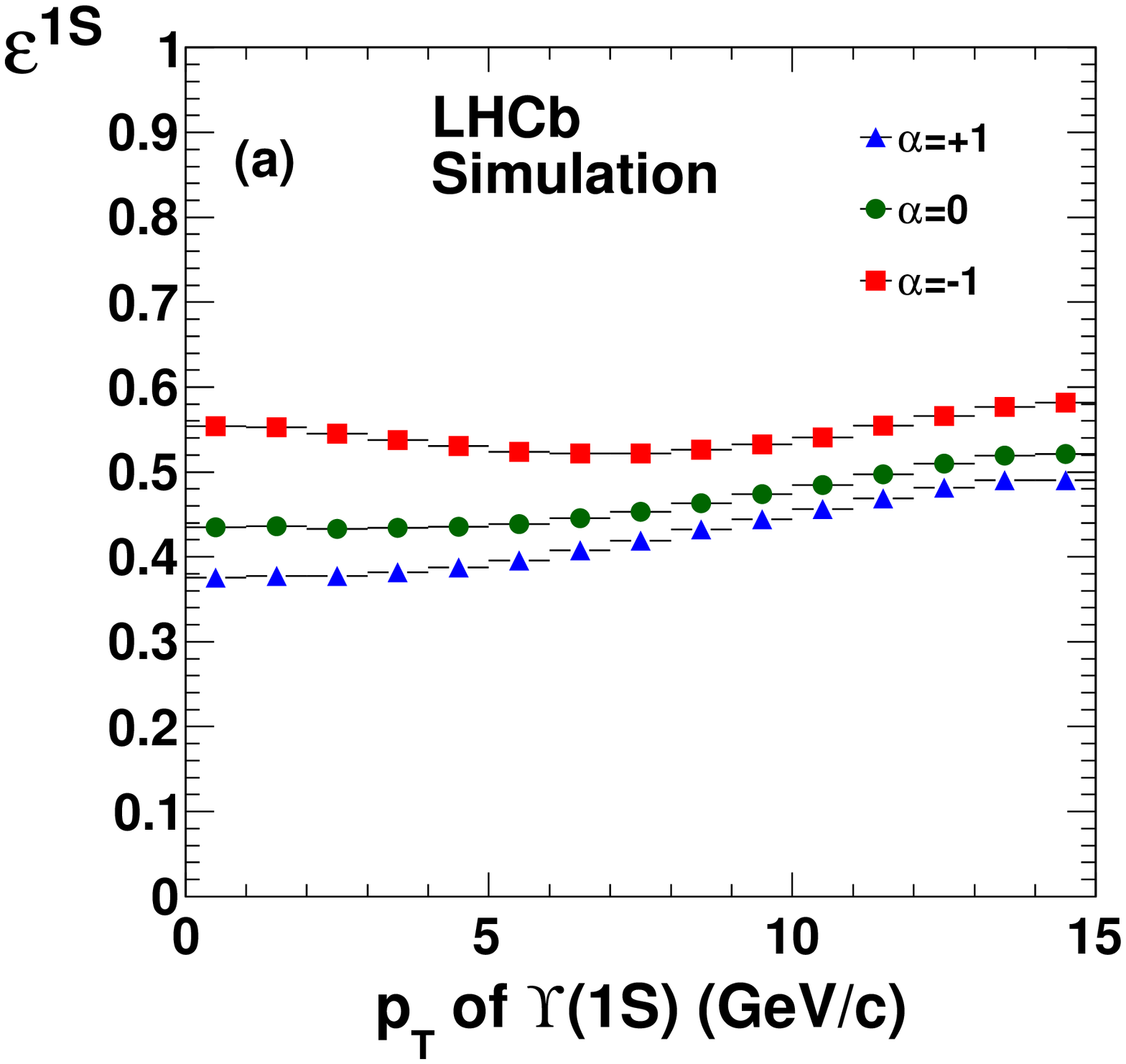}
 \includegraphics[width=0.45\textwidth, height=0.45\textwidth]{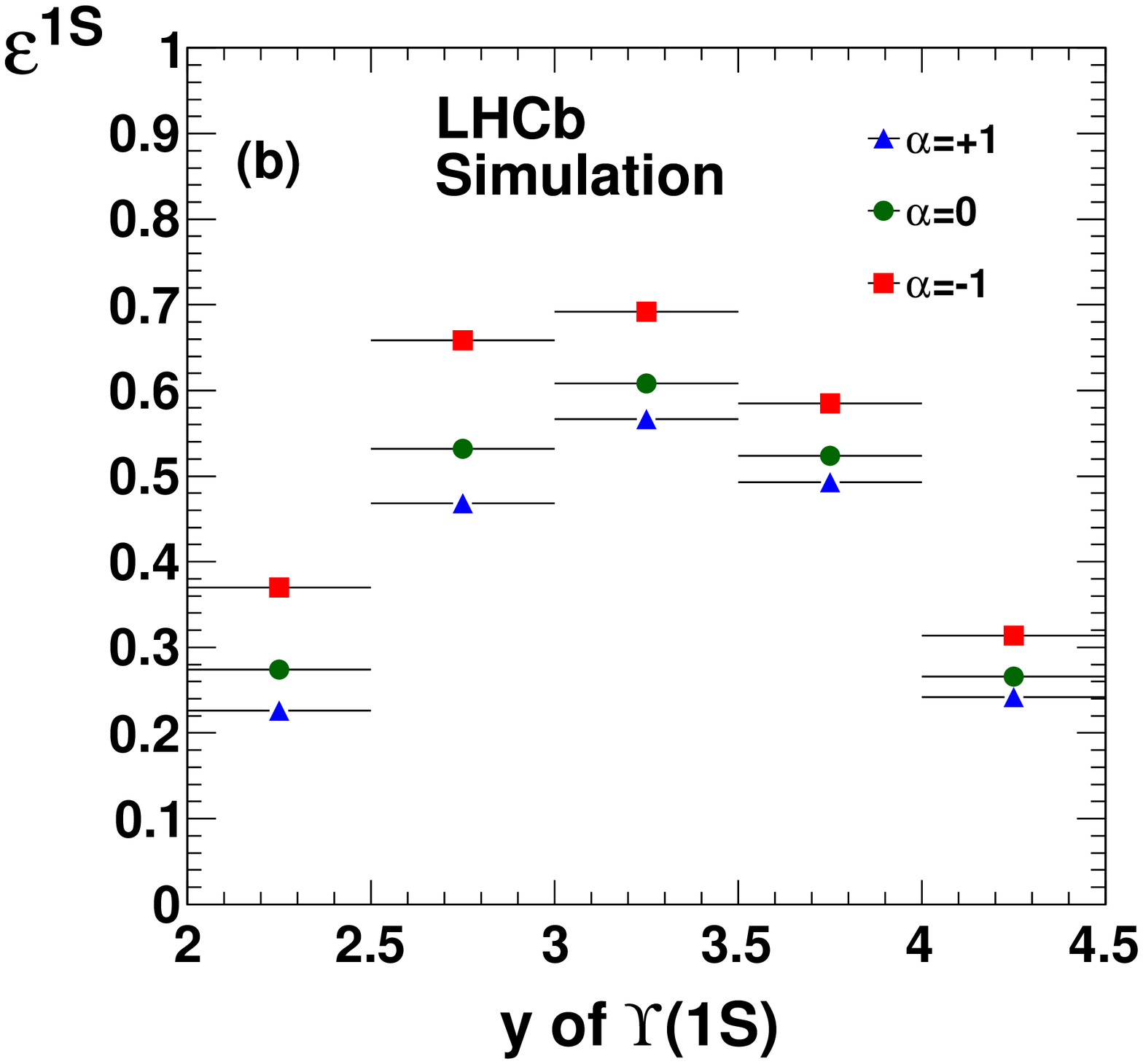}
 \caption{ \small{Total efficiency $\varepsilon$ of the $\ones$ 
as a function of (a) the $\ones$ transverse momentum  and 
(b) rapidity, estimated using 
the Monte Carlo simulation, for three different $\ones$ polarisation scenarios,
indicated by the parameter $\alpha$ described in the text.}}
 \label{fig::etot}
\ece
\end{figure}

Extensive  studies on dimuon decays~\cite{jpsi,cms,atlas}
have shown that the total efficiency
depends strongly on 
the initial polarisation state of the vector meson.
In this analysis, 
the influence of the unknown polarisation is studied 
in the helicity frame~\cite{helicity}  
using Monte Carlo simulation.
The angular distribution of the muons from the $\ups$, ignoring the azimuthal part, is 
\begin{eqnarray}
\frac{{\rm d}N}{{\rm d} \cos\theta} \, = \, \frac{1+\alpha \cos^2\theta}{2+2\alpha/3},
\end{eqnarray}
where $\theta$ is the angle between the direction of the $\mu^+$ momentum in the
$\ups$ centre-of-mass frame and  the direction of the $\ups$ momentum in the colliding proton
centre-of-mass frame. 
The values $\alpha=+1,-1,0$ correspond to  fully transverse, fully longitudinal, and no polarisation respectively.
Figure~\ref{fig::etot}  shows the $\ones$ total efficiency 
for these three scenarios, and  indicates that the polarisation significantly 
affects the efficiencies and that the effect depends on $p_{\rm T}$ and $y$. 
A similar behaviour is  observed for the $\twos$ and $\threes$ efficiencies.
Following this observation, in each $(\pt,y)$ bin 
the maximal difference between the polarised scenarios ($\alpha = \pm 1$) 
and the 
unpolarised scenario ($\alpha=0$) is taken as a systematic uncertainty on the efficiency.
This results in an uncertainty of up to $17\%$  on the 
integrated cross-sections and of up to 40\% in the individual bins.
Several other sources of possible systematic effects were studied.
They are summarised in Table~\ref{tab::allsys}.
\begin{table}[!th]
\begin{center}
\caption{\small{
Summary of the relative systematic uncertainties on the cross-section measurements. Ranges 
indicate variations depending on the ($\pt,y$) bin and the $\ups$ state.
All uncertainties are fully correlated among the bins.
}}\label{tab::allsys}
\vspace{0.4cm}
\begin{tabular}{l|c}

Source & Uncertainty (\%)\\\hline
Unknown $\ups$ polarisation         &    0.3--41.0\\ 
Trigger                             &    3.0  \\
Track reconstruction                &    2.4  \\
Track quality requirement           &    0.5  \\
Vertexing requirement               &    1.0  \\
Muon identification                 &    1.1  \\
Global event selection requirements &    0.6  \\
$\pt$ binning effect                &    1.0  \\
Fit function                        &    1.1--2.1\\
Luminosity                          &    3.5  \\
\end{tabular}
\end{center}
\end{table}

The trigger efficiency is determined on data using an unbiased sample of events 
that would trigger if the $\ups$ candidate were removed. The efficiency 
obtained with this method is compared with  the efficiency determined in the simulation.
The difference of 3.0\% is assigned as a
systematic uncertainty.

The uncertainty on the muon track reconstruction efficiency 
has been estimated 
 using a data driven tag-and-probe approach based on partially 
reconstructed $\jpsi\ra\mu^+\mu^-$ decays~\cite{tracking}, 
and found to be 2.4\% per muon pair.
Additional uncertainties 
are assigned, which account for the 
different behaviour in data and simulation of the 
track and vertex quality requirements.
The muon identification efficiency is 
measured 
using a tag-and-probe approach, which gives an uncertainty
on the efficiency of 1.1\%~\cite{jpsi}.

The measurement
of the global event selection 
efficiency
is taken as 
an additional uncertainty associated with the trigger.
An  uncertainty of 1.0\% is considered to account for the 
difference in the $\pt$ spectra in data and Monte Carlo simulation for the 
three $\ups$ states, which might have an effect on the 
correct bin assignment (``binning effect'').

The influence of the choice of the fit function describing 
the shape of the invariant mass distribution includes two 
components. The uncertainty on the shape of the background
distribution  is estimated 
using a different fit model (1.0--1.5\%).
The systematic associated with fixing the parameters
of the Crystal Ball function is estimated by varying
the central values within the parameters uncertainties, 
obtained when leaving them free to vary in the fit (0.5--1.4\%).

\section{Results}

\begin{figure}[h!]
\bce
 \includegraphics[height=0.28\textheight]{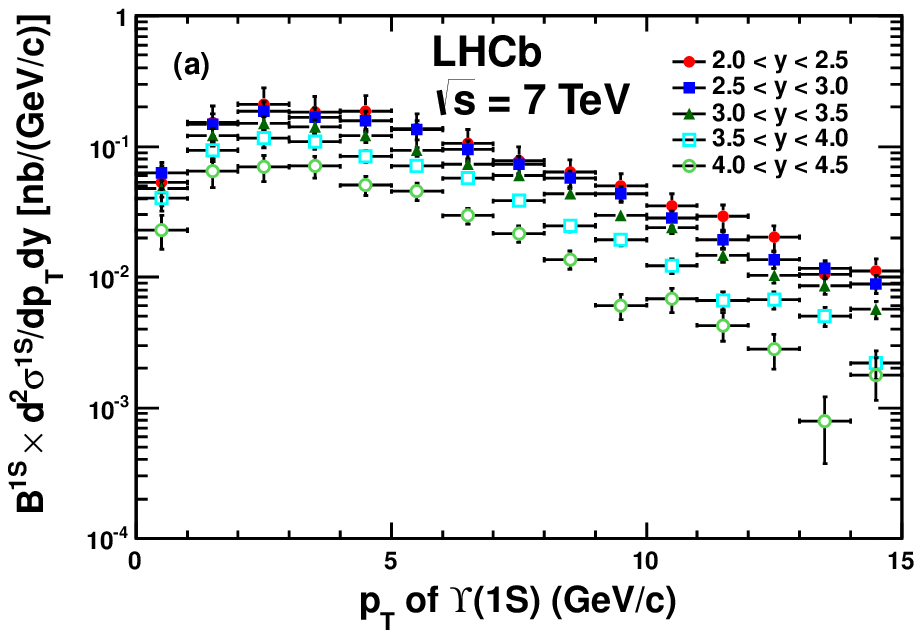}
 \includegraphics[height=0.28\textheight]{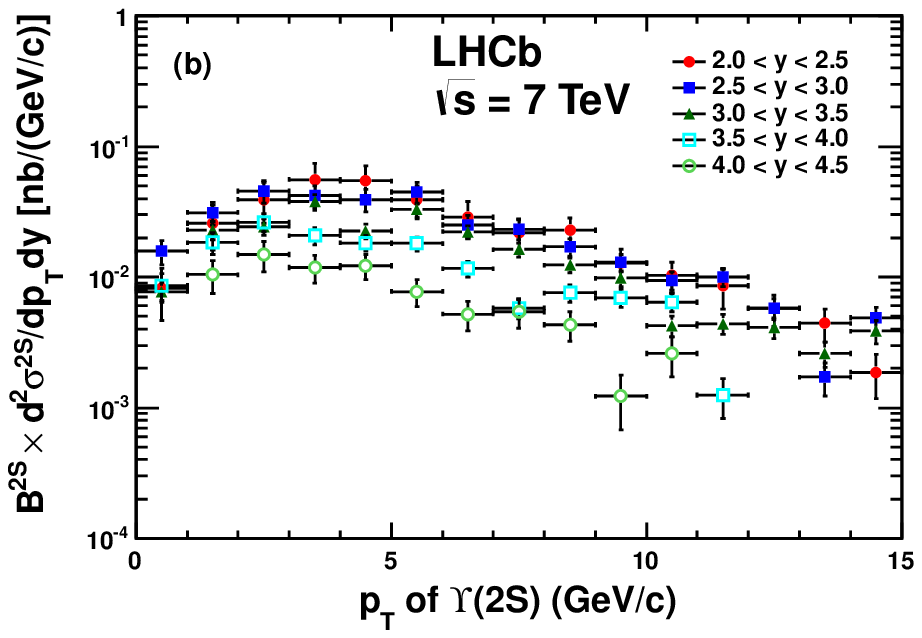}
 \includegraphics[height=0.28\textheight]{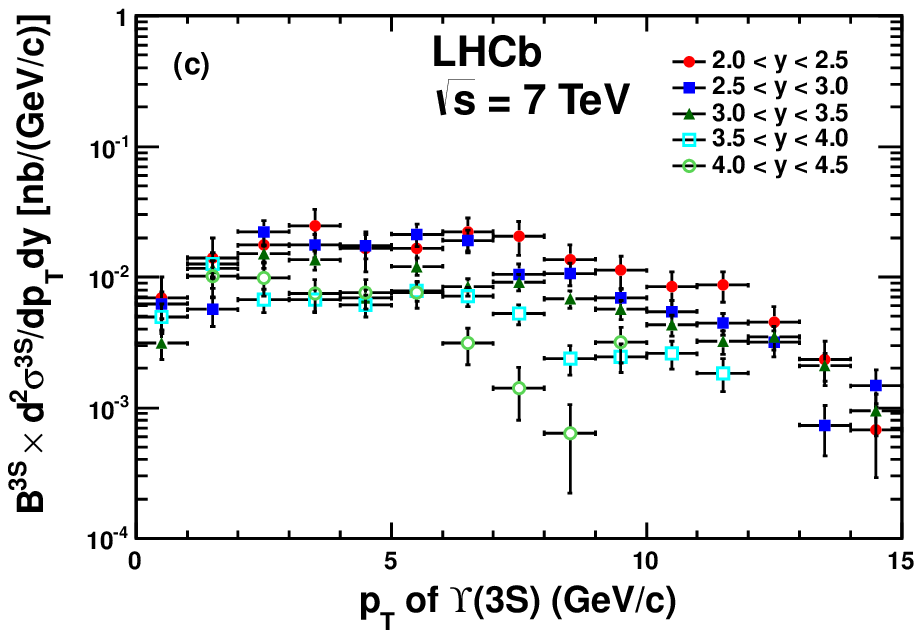}
 \caption{\small{Double differential $\upsmm$ cross-sections times dimuon branching
fractions as a function of $p_{\rm T}$ in bins of rapidity for (a) the $\ones$, (b) the 
$\twos$ and (c) the $\threes$. The error bars correspond to the total uncertainty 
 for each bin.}}\label{fig::xsec}
\ece
\end{figure}

The double differential cross-sections as a function of $\pt$ and $y$ are shown in Fig.~\ref{fig::xsec} and 
Tables~\ref{tab::xsec1s_corr}-\ref{tab::xsec3s_corr}.
The integrated cross-sections times branching fractions 
in the range $\pt<15\gevc$ and $2.0~<~y~<~4.5$ 
are measured to be
\begin{alignat*}{1}
\sigma(pp\ra \ones\,   X)\times\mathcal{B}^{1S} = 2.29\phantom{0}  \pm  0.01\phantom{0}   \pm 0.10\phantom{0}\,\, _{-0.37}^{+0.19}~{\rm nb}, \\
\sigma(pp\ra \twos\,   X)\times\mathcal{B}^{2S} = 0.562  \pm  0.007   \pm 0.023\, _{-0.092}^{+0.048}~{\rm nb}, \\
\sigma(pp\ra \threes\, X)\times\mathcal{B}^{3S} = 0.283  \pm  0.005 \pm 0.012\, _{-0.048}^{+0.025}~{\rm nb},
\end{alignat*}
where the first uncertainties
are statistical, the second systematic and the third are due to the unknown polarisation of the 
three $\ups$ states.
The integrated $\ones$ cross-section is about a factor one hundred smaller than the integrated $\jpsi$ cross-section
in the identical region of $\pt$ and $y$~\cite{jpsi}, and a factor three smaller than the 
integrated $\ones$ cross-section in the central region, as measured by CMS~\cite{cms}
and ATLAS~\cite{atlas}. 

\begin{figure}[htb!]
\bce
  \includegraphics[height=0.28\textheight]{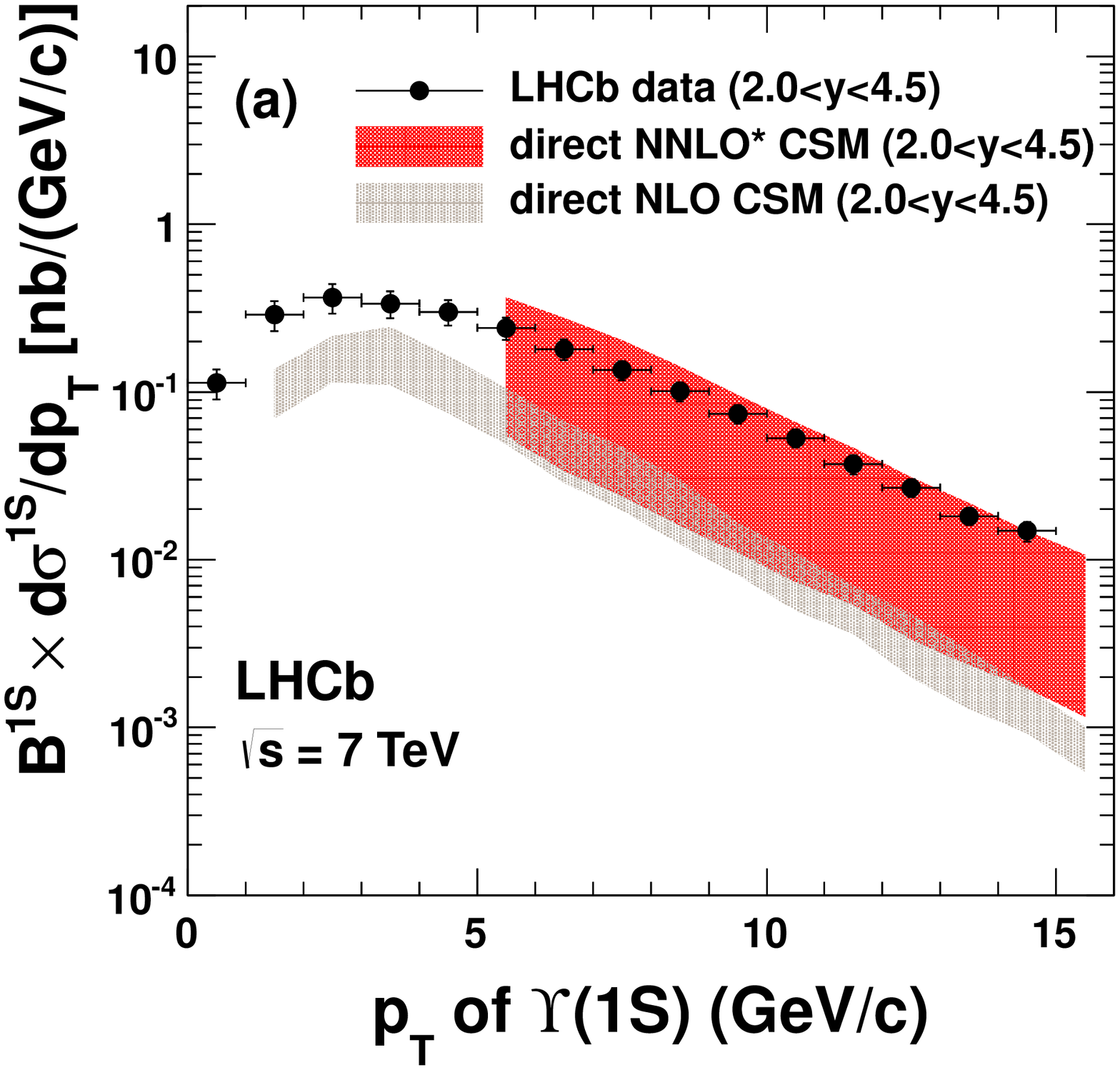}
  \includegraphics[height=0.28\textheight]{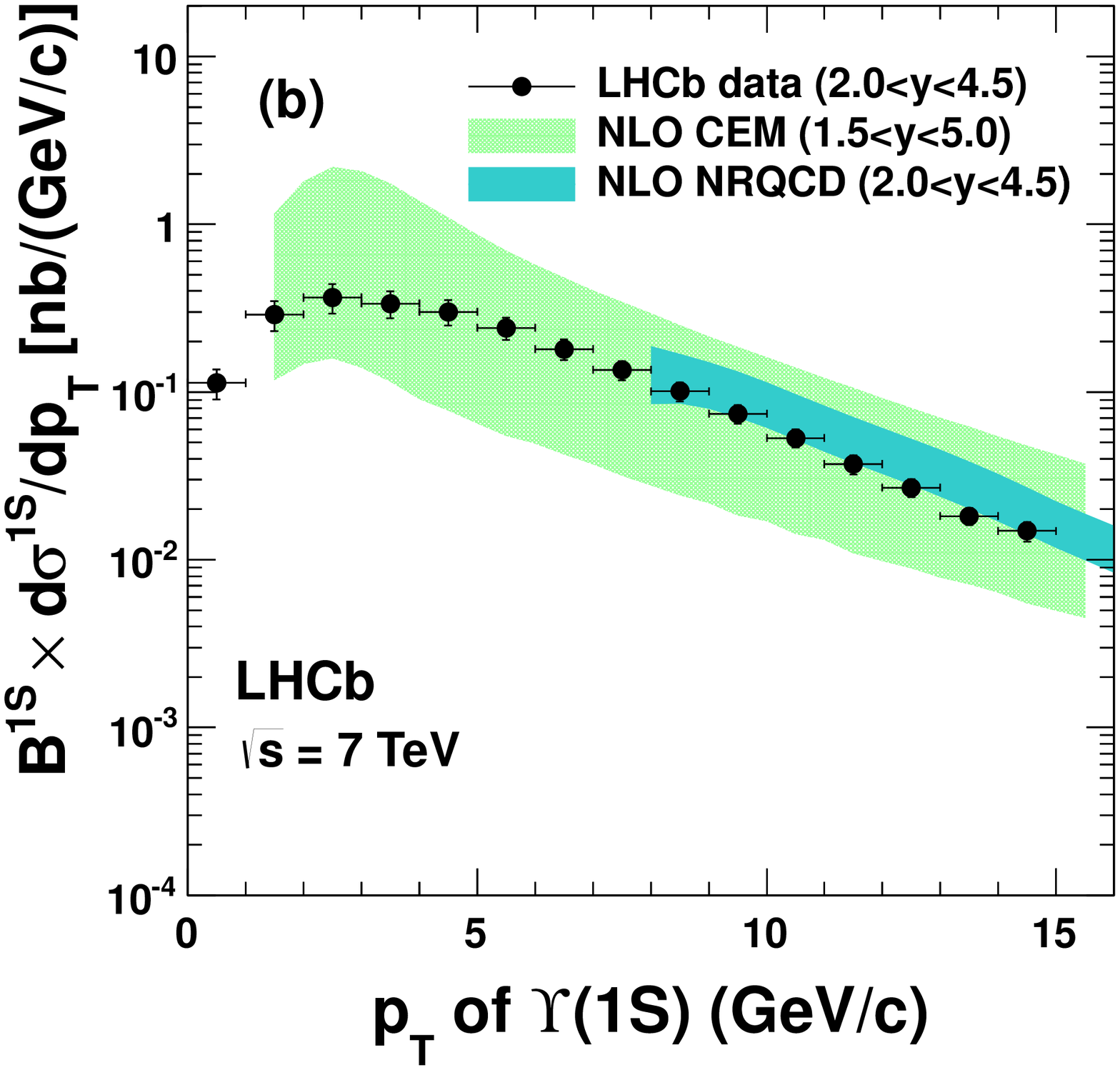}
 \caption{\small{ Differential $\onesmm$ production cross-section times dimuon branching
fraction  as a function of $p_{\rm T}$ 
integrated over $y$ in the range 2.0--4.5, compared with the predictions from (a) the 
NNLO* CSM~\cite{jp} for 
direct production, and (b) the 
NLO NRQCD~\cite{chao} and CEM~\cite{CEM}. The error bars on the data correspond to the total 
uncertainties for each bin, while the bands indicate the uncertainty on the theory prediction.}}
\label{fig::1stheory}
\ece
\end{figure}

\begin{figure}[htb!]
\vspace{1cm}
\bce
  \includegraphics[height=0.28\textheight]{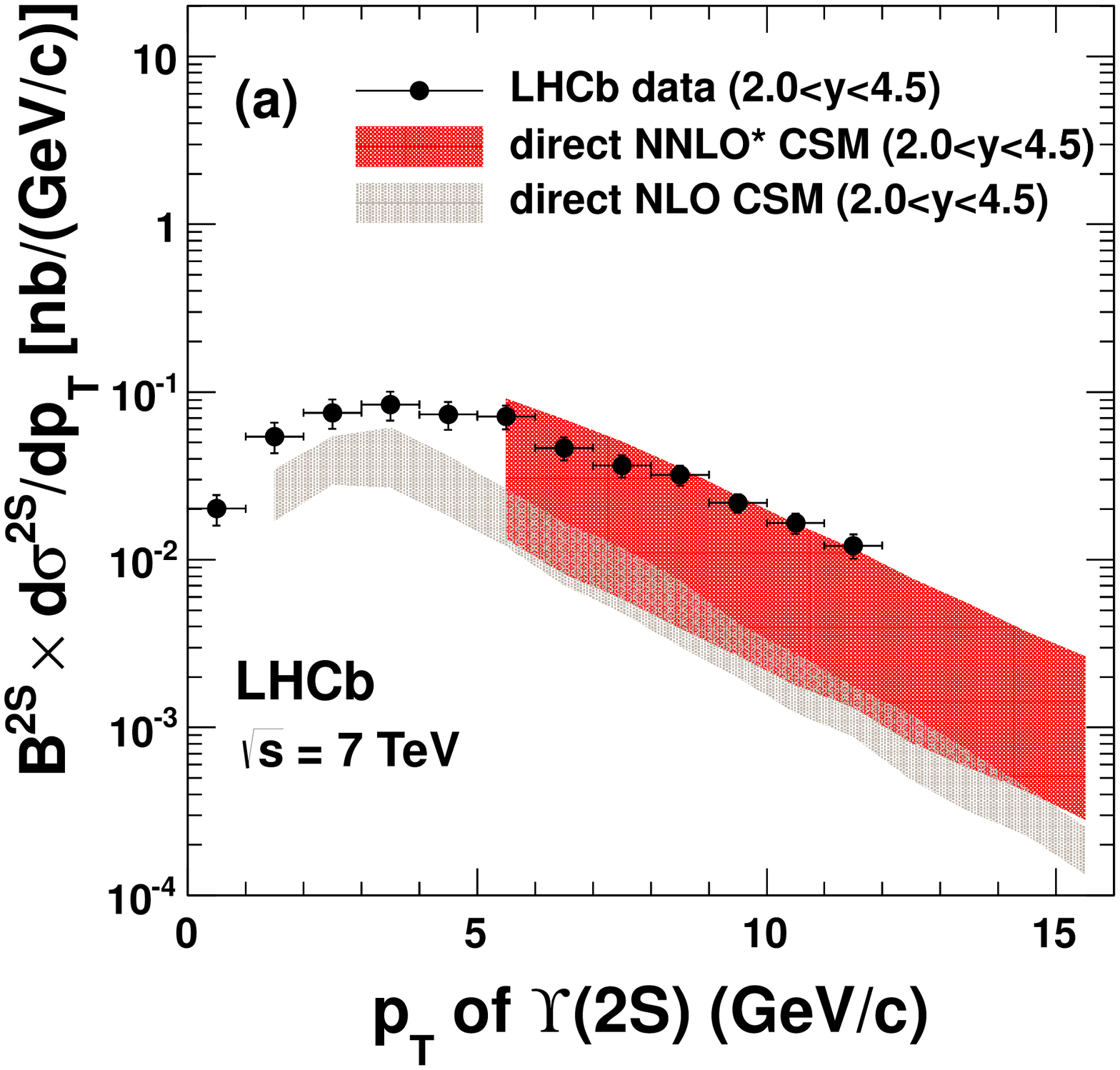}
  \includegraphics[height=0.28\textheight]{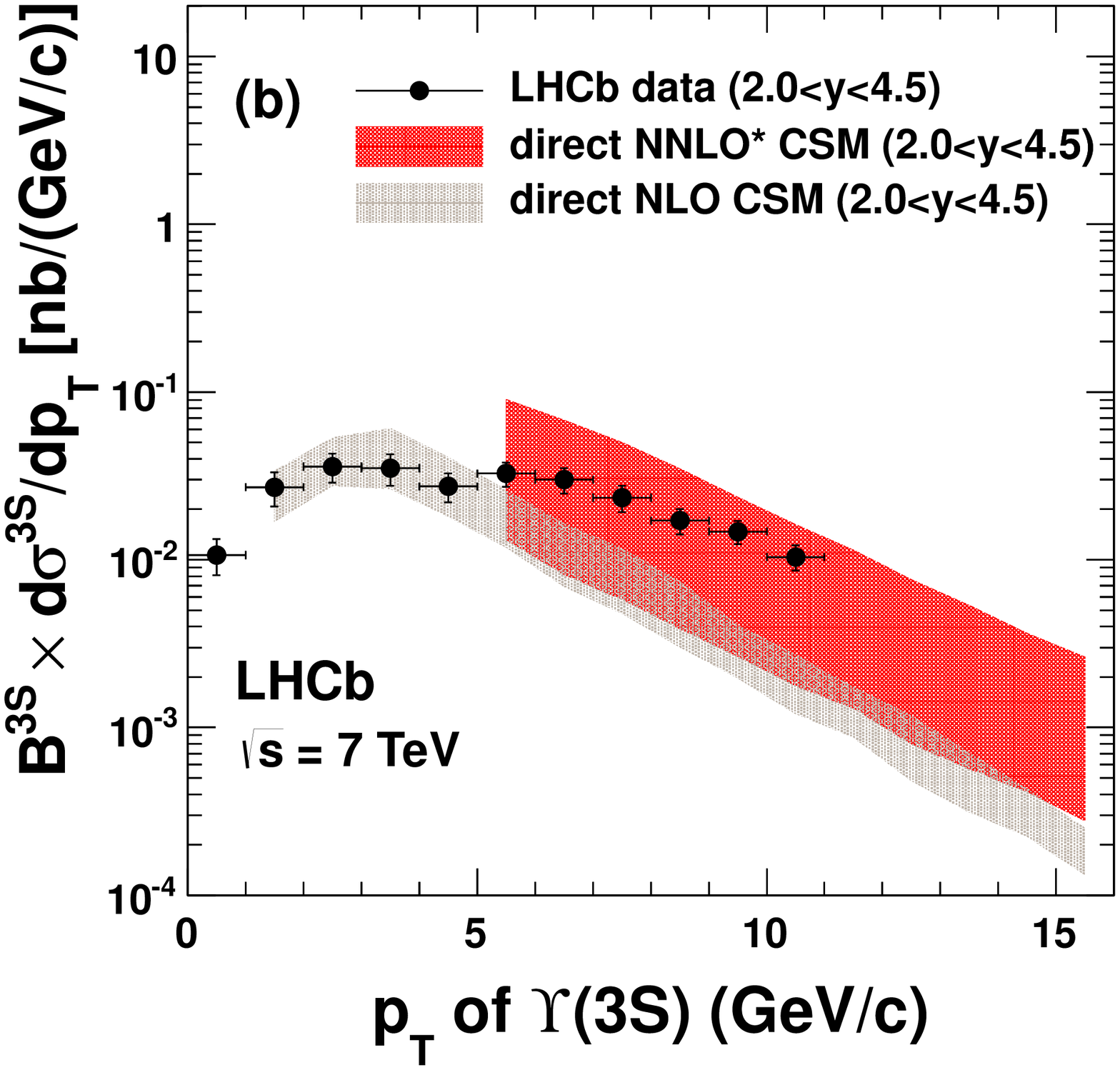}
 \caption{\small{Differential (a) $\twosmm$ and (b) $\threesmm$ 
production cross-sections times dimuon branching fractions  
as a function of $p_{\rm T}$ 
integrated over $y$ in the range 2.0--4.5, compared with
the predictions from the NNLO* CSM for
direct production~\cite{jp}. The error bars on the data correspond to the total 
uncertainties for each bin, while the bands indicate the uncertainty on the theory prediction.}}
\label{fig::2stheory}
\ece
\end{figure}

\begin{figure}[htb!]
\bce
 \includegraphics[ height=0.28\textheight]{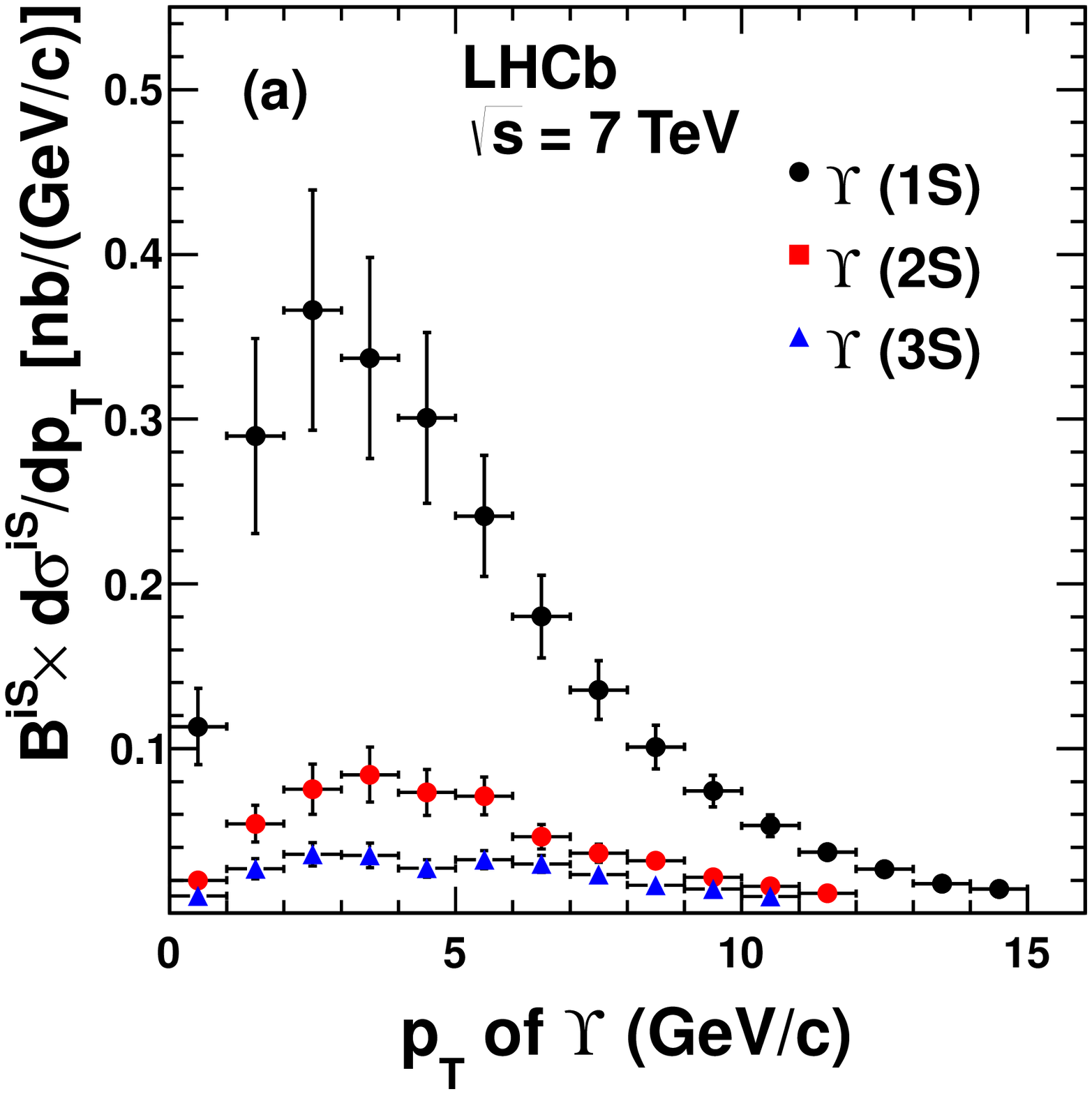}
 \includegraphics[ height=0.28\textheight]{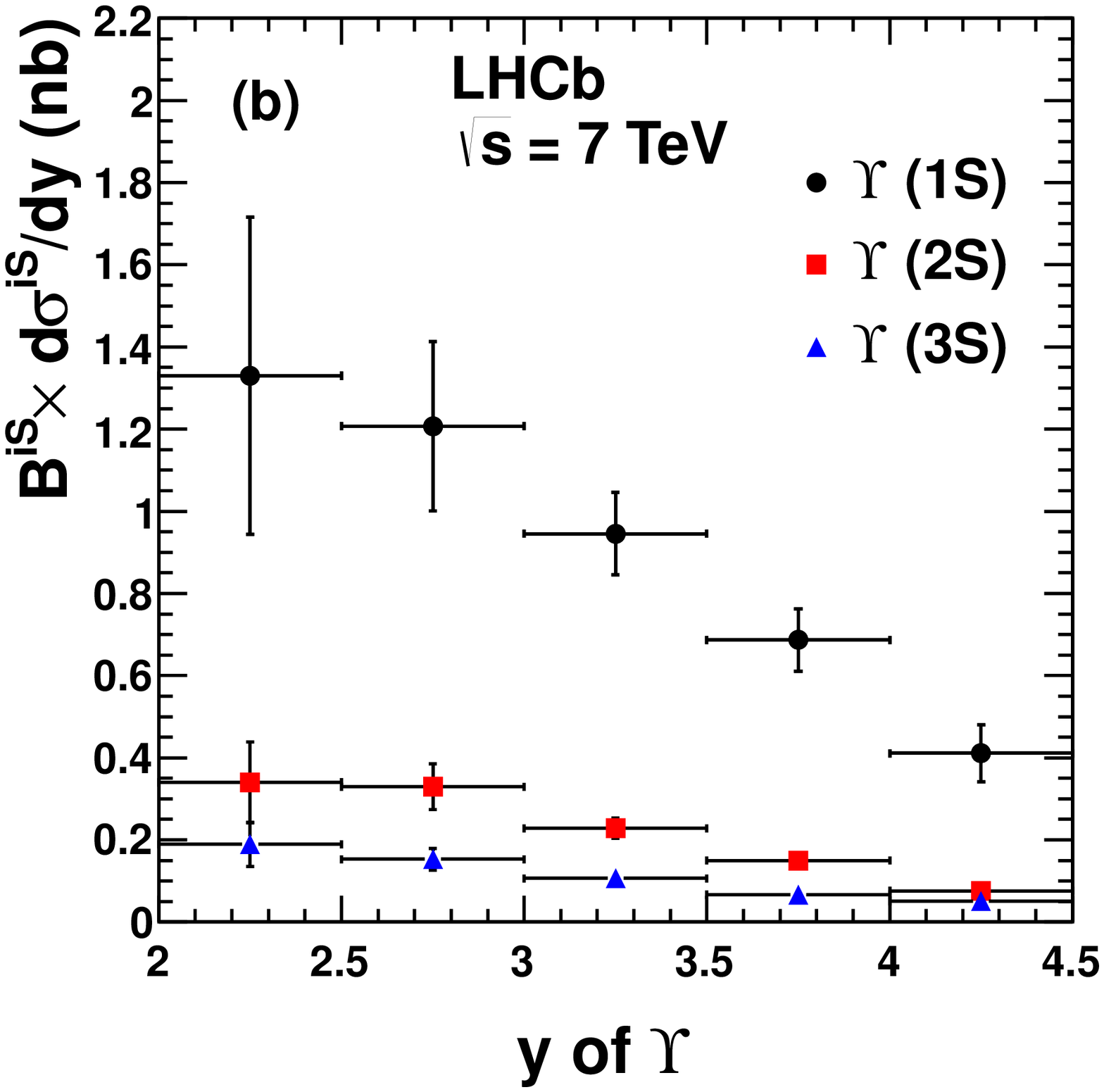}
 \caption{\small{ Differential cross-sections of $\ones,\twos$ and $\threes$ 
times dimuon 
branching fractions as a function of (a) 
 $\pt$  integrated over $y$  and (b)  $y$ integrated 
over $\pt$. The error bars on the data correspond to the total 
uncertainties for each bin.}
}\label{fig::allups}
\ece
\end{figure}

\begin{figure}[htb!]
\bce
 \includegraphics[height=0.28\textheight]{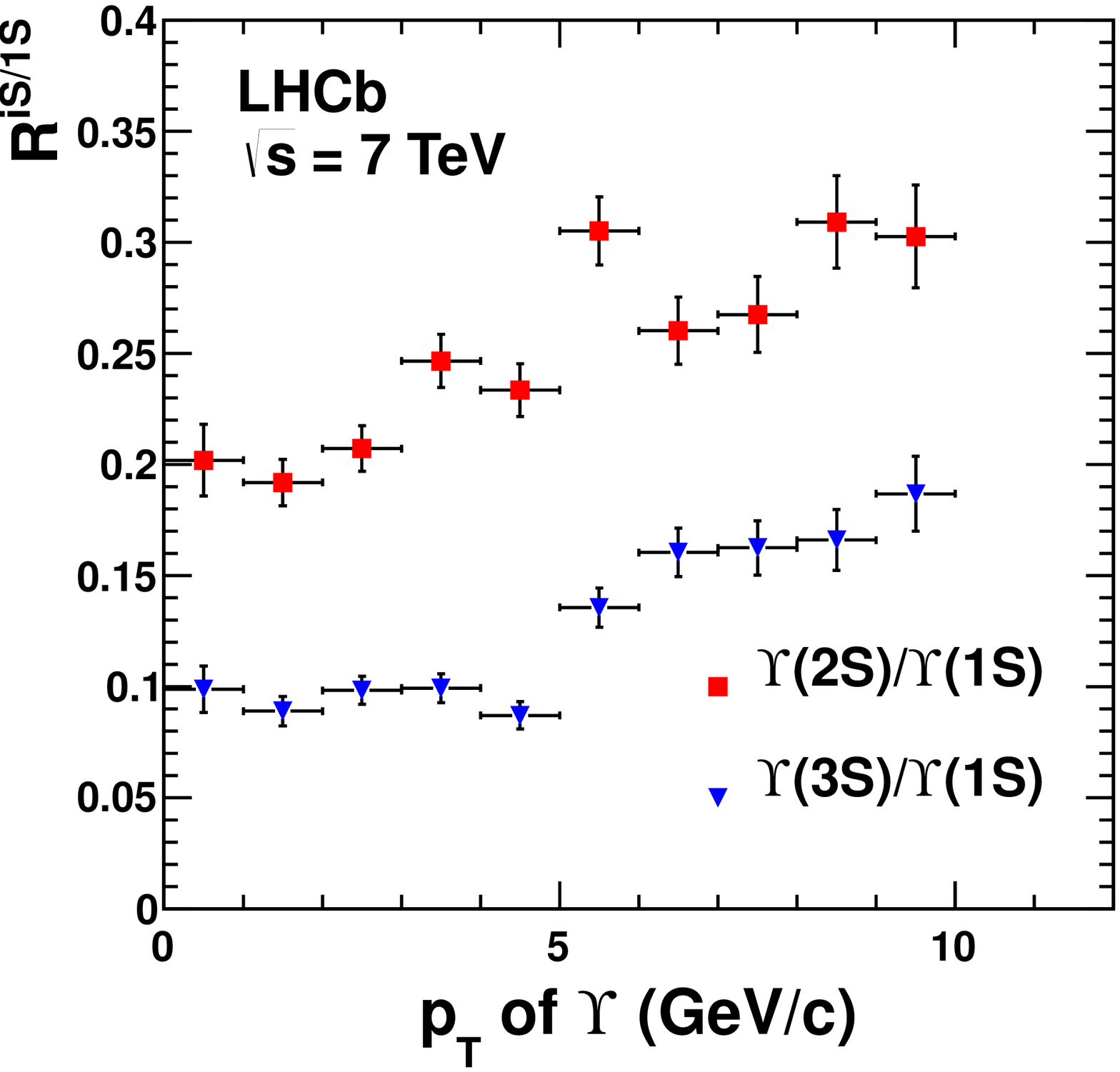}
 \caption{ 
\small{Ratios of $\twosmm$ and $\threesmm$ 
 with respect to $\onesmm$
 as a function of $\pt$ of the $\ups$ in the
range $2.0<y<4.5$, assuming no polarisation. 
The error bars on the data correspond to the total 
uncertainties for each bin except for that due to the unknown polarisation,
which ranges between 15\% and 26\% as listed in Table~\ref{tab::ratios}.
}}\label{fig::myratios}
\ece
\end{figure}

Figure~\ref{fig::1stheory} compares the LHCb measurement of the differential $\onesmm$ production 
cross-section 
with several theory predictions
in the LHCb acceptance region.
In Fig.~\ref{fig::1stheory}(a) the data are compared to 
direct production as calculated from a NNLO* colour-singlet model~\cite{jp,jp2}, 
where 
the notation NNLO* denotes an evaluation that is not a complete next-to-next 
leading order computation and that can be 
affected by logarithmic corrections, which are not easily quantifiable. 
Direct production as calculated from NLO CSM is also represented.\label{jptext}
In Fig.~\ref{fig::1stheory}(b) the data are compared to two model 
predictions for the $\ones$ production: the calculation from NRQCD at NLO, 
including contributions from $\chi_b$ and higher $\ups$ states  decays, summing the colour-singlet and colour-octet 
contributions~\cite{chao}, and 
the calculation  from the NLO CEM, including 
contributions from $\chi_b$ and higher $\ups$ states  decays~\cite{CEM}.
Note that the NNLO* theoretical model computes the direct 
$\ones$ production, whereas the LHCb 
measurement includes $\ones$ from $\chi_b$, $\twos$ and $\threes$ decays. 
However, taking into
account the feed-down contribution, which has been measured
 to be of the order of 50\%~\cite{cdf_chib},
a satisfactory agreement is found with the theoretical predictions.
Figure~\ref{fig::2stheory} 
compares the LHCb measurement of the differential $\twos$ and $\threes$ 
production cross-sections times branching fraction
with the NNLO* theory predictions of direct production.
It can be seen that the agreement with the theory is better 
for the $\threes$, which is expected to be less affected by 
feed-down. 
At present there is no measurement of the contribution of 
feed-down to the $\twos$ and 
$\threes$ inclusive rate.
The cross-sections times the dimuon branching fractions  for the three $\ups$ states are compared 
in Fig.~\ref{fig::allups} as a function of rapidity and transverse momentum.
The cross-section results are used to evaluate 
the ratios $R^{iS/1S}$ of the $\twos$ to $\ones$ and $\threes$ to $\ones$
cross-sections times the dimuon branching fractions.
Most of the systematic uncertainties on the cross-sections cancel in the 
ratio,  except those 
due to the size of the data sample, the choice of fit function and  the 
unknown polarisation of the different states.
The polarisation uncertainty has been evaluated for the scenarios 
in which one of the two $\ups$ states is completely polarised (either  
transversely or longitudinally) and the other is not polarised. The 
maximum difference of these two cases 
ranges between 15\% and 26\%.
The ratios $R^{iS/1S}, i=2,3,$ 
are given in Table~\ref{tab::ratios} and 
shown in Fig.~\ref{fig::myratios}.
The polarisation uncertainty is not included in these figures.
The results agree well with the corresponding ratio measurements from CMS~\cite{cms} 
in the $\pt$ range common to both experiments.

\section{Conclusions}

The differential cross-sections $\ismm$, for $i=1,2,3$,
are measured as a function of 
the $\ups$ transverse momentum and rapidity in the region $\pt<15~\gevc$, 2.0 $< y <$ 4.5 in the 
LHCb experiment. The analysis 
is based on a data sample corresponding to an integrated luminosity of 25~$\invpb$ collected at the Large Hadron Collider 
at a centre-of-mass energy of $\sqrt{s} = 7 ~\tev$.
The results obtained are compatible with previous measurements in $pp$ collisions 
at the same centre-of-mass energy, performed by ATLAS and CMS in a different region of rapidity~\cite{cms,atlas}. 
This is the first measurement of $\ups$ production in the forward region at $\sqrt{s} = 7~\tev$.
A comparison with theoretical models shows good  agreement with the measured $\ups$ cross-sections.
The measurement of the differential cross-sections  is 
not sufficient  to discriminate amongst the various models, and studies of other observables such as 
the $\ups$ polarisations will be necessary.

\section{Acknowledgements}

We thank P.~Artoisenet, M.~Butensch\"{o}n, K.-T.~Chao, B.~Kniehl, J.-P.~Lansberg and R.~Vogt for 
providing theoretical predictions of \ups\ cross-sections in the LHCb acceptance range.
\noindent We express our gratitude to our colleagues in the CERN accelerator
departments for the excellent performance of the LHC. We thank the
technical and administrative staff at CERN and at the LHCb institutes,
and acknowledge support from the National Agencies: CAPES, CNPq,
FAPERJ and FINEP (Brazil); CERN; NSFC (China); CNRS/IN2P3 (France);
BMBF, DFG, HGF and MPG (Germany); SFI (Ireland); INFN (Italy); FOM and
NWO (The Netherlands); SCSR (Poland); ANCS (Romania); MinES of Russia and
Rosatom (Russia); MICINN, XuntaGal and GENCAT (Spain); SNSF and SER
(Switzerland); NAS Ukraine (Ukraine); STFC (United Kingdom); NSF
(USA). We also acknowledge the support received from the ERC under FP7
and the Region Auvergne.

\clearpage
\renewcommand{\arraystretch}{1.6}
 \begin{table}[!ht] 
\begin{center} 
\caption{\small{Double differential cross-section  $\onesmm$
    as a function of rapidity  and transverse momentum, in pb/(\gevc). The first uncertainty is statistical,
    the second is systematic, and the third is due to the unknown polarisation of the $\ones$.}
}\label{tab::xsec1s_corr} 
\resizebox{\textwidth}{!}{ 
\begin{tabular}{|c|c|c|c|c|c|} 
\hline 
 $\pt$ & $2.0<y<2.5$ & $2.5<y<3.0$ & $3.0<y<3.5$ & $3.5<y<4.0$ & $4.0<y<4.5$  \\ 
 \small{(\gevc)} &  &  &  &  &  \\ \hline
   0--1  &    53.1 $\pm$     4.0 $\pm$     2.5 $_{-   17.3}^{+    8.9}$  &    62.6 $\pm$     3.0 $\pm$     2.9 $_{-   11.5}^{+    6.1}$  &    48.0 $\pm$     2.4 $\pm$     2.2 $_{-    5.8}^{+    3.1}$  &    40.1 $\pm$     2.4 $\pm$     1.9 $_{-    7.0}^{+    3.9}$  &    22.9 $\pm$     2.7 $\pm$     1.1 $_{-    5.9}^{+    3.4}$ \\\hline
   1--2  &   152.5 $\pm$     6.8 $\pm$     7.2 $_{-   50.4}^{+   25.7}$  &   148.8 $\pm$     4.7 $\pm$     7.0 $_{-   27.5}^{+   14.6}$  &   120.5 $\pm$     3.8 $\pm$     5.6 $_{-   14.0}^{+    7.5}$  &    93.3 $\pm$     3.7 $\pm$     4.3 $_{-   14.8}^{+    8.1}$  &    64.5 $\pm$     4.5 $\pm$     3.0 $_{-   15.0}^{+    8.7}$ \\\hline
   2--3  &   211.0 $\pm$     8.0 $\pm$    10.0 $_{-   67.2}^{+   34.3}$  &   185.3 $\pm$     5.2 $\pm$     8.7 $_{-   34.4}^{+   18.1}$  &   150.0 $\pm$     4.3 $\pm$     7.0 $_{-   17.4}^{+    9.2}$  &   116.1 $\pm$     4.1 $\pm$     5.4 $_{-   15.5}^{+    8.4}$  &    69.8 $\pm$     4.6 $\pm$     3.3 $_{-   14.6}^{+    8.3}$ \\\hline
   3--4  &   184.3 $\pm$     7.3 $\pm$     8.8 $_{-   56.3}^{+   28.8}$  &   167.7 $\pm$     4.9 $\pm$     7.9 $_{-   29.3}^{+   15.6}$  &   141.9 $\pm$     4.2 $\pm$     6.6 $_{-   15.0}^{+    8.0}$  &   109.7 $\pm$     4.0 $\pm$     5.1 $_{-   11.9}^{+    6.3}$  &    70.6 $\pm$     4.6 $\pm$     3.3 $_{-   12.2}^{+    6.7}$ \\\hline
   4--5  &   187.3 $\pm$     7.3 $\pm$     8.9 $_{-   54.8}^{+   27.9}$  &   158.4 $\pm$     4.8 $\pm$     7.4 $_{-   26.4}^{+   14.0}$  &   120.9 $\pm$     3.9 $\pm$     5.7 $_{-   11.3}^{+    6.0}$  &    84.6 $\pm$     3.5 $\pm$     4.0 $_{-    7.0}^{+    3.7}$  &    50.4 $\pm$     3.8 $\pm$     2.4 $_{-    7.0}^{+    3.7}$ \\\hline
   5--6  &   138.0 $\pm$     6.2 $\pm$     6.6 $_{-   38.3}^{+   19.4}$  &   134.5 $\pm$     4.4 $\pm$     6.3 $_{-   20.8}^{+   11.0}$  &    94.2 $\pm$     3.5 $\pm$     4.4 $_{-    7.3}^{+    3.8}$  &    70.6 $\pm$     3.2 $\pm$     3.3 $_{-    4.0}^{+    2.1}$  &    45.3 $\pm$     3.6 $\pm$     2.1 $_{-    4.9}^{+    2.5}$ \\\hline
   6--7  &   105.3 $\pm$     5.3 $\pm$     5.0 $_{-   27.6}^{+   14.0}$  &    95.2 $\pm$     3.7 $\pm$     4.5 $_{-   13.7}^{+    7.2}$  &    73.5 $\pm$     3.0 $\pm$     3.5 $_{-    4.6}^{+    2.4}$  &    57.0 $\pm$     2.9 $\pm$     2.7 $_{-    1.9}^{+    1.0}$  &    29.5 $\pm$     2.8 $\pm$     1.4 $_{-    2.5}^{+    1.2}$ \\\hline
   7--8  &    78.3 $\pm$     4.5 $\pm$     3.7 $_{-   19.4}^{+    9.8}$  &    72.9 $\pm$     3.2 $\pm$     3.4 $_{-    9.6}^{+    5.0}$  &    60.2 $\pm$     2.7 $\pm$     2.8 $_{-    3.0}^{+    1.6}$  &    38.3 $\pm$     2.3 $\pm$     1.8 $_{-    0.8}^{+    0.4}$  &    21.6 $\pm$     2.4 $\pm$     1.0 $_{-    1.5}^{+    0.7}$ \\\hline
   8--9  &    63.5 $\pm$     4.0 $\pm$     3.0 $_{-   14.8}^{+    7.5}$  &    57.0 $\pm$     2.8 $\pm$     2.7 $_{-    6.8}^{+    3.6}$  &    43.3 $\pm$     2.3 $\pm$     2.0 $_{-    1.9}^{+    1.0}$  &    24.7 $\pm$     1.9 $\pm$     1.2 $_{-    0.6}^{+    0.3}$  &    13.6 $\pm$     1.9 $\pm$     0.6 $_{-    0.8}^{+    0.4}$ \\\hline
   9--10  &    50.1 $\pm$     3.5 $\pm$     2.4 $_{-   10.8}^{+    5.5}$  &    43.2 $\pm$     2.4 $\pm$     2.0 $_{-    5.0}^{+    2.6}$  &    29.8 $\pm$     1.9 $\pm$     1.4 $_{-    1.0}^{+    0.5}$  &    19.4 $\pm$     1.6 $\pm$     0.9 $_{-    0.6}^{+    0.3}$  &     6.1 $\pm$     1.2 $\pm$     0.3 $_{-    0.3}^{+    0.1}$ \\\hline
  10--11  &    35.4 $\pm$     2.9 $\pm$     1.7 $_{-    7.3}^{+    3.7}$  &    28.2 $\pm$     1.9 $\pm$     1.3 $_{-    3.0}^{+    1.6}$  &    23.9 $\pm$     1.7 $\pm$     1.1 $_{-    0.8}^{+    0.4}$  &    12.3 $\pm$     1.3 $\pm$     0.6 $_{-    0.5}^{+    0.2}$  &     6.8 $\pm$     1.3 $\pm$     0.3 $_{-    0.4}^{+    0.2}$ \\\hline
  11--12  &    29.3 $\pm$     2.6 $\pm$     1.4 $_{-    5.8}^{+    2.9}$  &    19.4 $\pm$     1.6 $\pm$     0.9 $_{-    1.9}^{+    1.0}$  &    14.7 $\pm$     1.3 $\pm$     0.7 $_{-    0.6}^{+    0.3}$  &     6.7 $\pm$     0.9 $\pm$     0.3 $_{-    0.2}^{+    0.1}$  &     4.3 $\pm$     1.0 $\pm$     0.2 $_{-    0.3}^{+    0.1}$ \\\hline
  12--13  &    20.3 $\pm$     2.1 $\pm$     1.0 $_{-    3.7}^{+    1.9}$  &    13.7 $\pm$     1.3 $\pm$     0.6 $_{-    1.3}^{+    0.7}$  &    10.3 $\pm$     1.1 $\pm$     0.5 $_{-    0.3}^{+    0.2}$  &     6.7 $\pm$     0.9 $\pm$     0.3 $_{-    0.2}^{+    0.1}$  &     2.8 $\pm$     0.8 $\pm$     0.1 $_{-    0.2}^{+    0.1}$ \\\hline
  13--14  &    10.4 $\pm$     1.5 $\pm$     0.5 $_{-    1.9}^{+    0.9}$  &    11.6 $\pm$     1.2 $\pm$     0.5 $_{-    1.1}^{+    0.6}$  &     8.6 $\pm$     1.0 $\pm$     0.4 $_{-    0.2}^{+    0.1}$  &     5.0 $\pm$     0.8 $\pm$     0.2 $_{-    0.2}^{+    0.1}$  &     0.8 $\pm$     0.4 $\pm$     0.0 $_{-    0.1}^{+    0.0}$ \\\hline
  14--15  &    11.2 $\pm$     1.5 $\pm$     0.5 $_{-    2.0}^{+    1.0}$  &     8.9 $\pm$     1.0 $\pm$     0.4 $_{-    0.8}^{+    0.4}$  &     5.7 $\pm$     0.8 $\pm$     0.3 $_{-    0.2}^{+    0.1}$  &     2.2 $\pm$     0.5 $\pm$     0.1 $_{-    0.1}^{+    0.0}$  &     1.8 $\pm$     0.6 $\pm$     0.1 $_{-    0.1}^{+    0.1}$ \\\hline
\end{tabular}
}
\end{center} 
\end{table}

 \begin{table}[!ht] 
\begin{center} 
\caption{\small{Double differential cross-section  $\twosmm$
    as a function of rapidity  and transverse momentum, in pb/(\gevc). 
    The first uncertainty is statistical,
    the second is systematic, and the third is due to the unknown polarisation of the $\twos$. 
     Regions where the number of events was not sufficient to perform a measurement are
     indicated with a dash.}
}\label{tab::xsec2s_corr} 
\resizebox{\textwidth}{!}{ 
\begin{tabular}{|c|c|c|c|c|c|} 
\hline 
 $\pt$ & $2.0<y<2.5$ & $2.5<y<3.0$ & $3.0<y<3.5$ & $3.5<y<4.0$ & $4.0<y<4.5$  \\ 
 \small{(\gevc)} &  & &  &  &  \\ \hline
   0--1  &     8.2 $\pm$     1.7 $\pm$     0.4 $_{-    3.1}^{+    1.5}$  &    15.8 $\pm$     1.6 $\pm$     0.7 $_{-    2.8}^{+    1.5}$  &     7.8 $\pm$     1.0 $\pm$     0.4 $_{-    0.8}^{+    0.4}$  &     8.6 $\pm$     1.2 $\pm$     0.4 $_{-    1.5}^{+    0.8}$  & - \\\hline
   1--2  &    25.8 $\pm$     2.9 $\pm$     1.2 $_{-    9.2}^{+    4.6}$  &    31.2 $\pm$     2.2 $\pm$     1.5 $_{-    5.6}^{+    3.1}$  &    23.0 $\pm$     1.7 $\pm$     1.1 $_{-    2.9}^{+    1.6}$  &    18.3 $\pm$     1.6 $\pm$     0.9 $_{-    2.8}^{+    1.6}$  &    10.4 $\pm$     1.8 $\pm$     0.5 $_{-    2.3}^{+    1.4}$ \\\hline
   2--3  &    39.3 $\pm$     3.6 $\pm$     1.9 $_{-   12.9}^{+    6.4}$  &    45.7 $\pm$     2.6 $\pm$     2.1 $_{-    8.2}^{+    4.5}$  &    24.4 $\pm$     1.8 $\pm$     1.1 $_{-    2.9}^{+    1.5}$  &    26.3 $\pm$     2.0 $\pm$     1.2 $_{-    3.4}^{+    1.9}$  &    14.9 $\pm$     2.2 $\pm$     0.7 $_{-    3.2}^{+    1.8}$ \\\hline
   3--4  &    55.8 $\pm$     4.2 $\pm$     2.6 $_{-   17.4}^{+    8.9}$  &    42.1 $\pm$     2.5 $\pm$     2.0 $_{-    7.3}^{+    3.8}$  &    37.8 $\pm$     2.2 $\pm$     1.8 $_{-    4.3}^{+    2.2}$  &    20.8 $\pm$     1.8 $\pm$     1.0 $_{-    2.4}^{+    1.3}$  &    11.9 $\pm$     1.9 $\pm$     0.6 $_{-    2.1}^{+    1.2}$ \\\hline
   4--5  &    54.5 $\pm$     4.1 $\pm$     2.6 $_{-   15.9}^{+    8.2}$  &    39.2 $\pm$     2.4 $\pm$     1.8 $_{-    6.7}^{+    3.6}$  &    22.6 $\pm$     1.7 $\pm$     1.1 $_{-    2.0}^{+    1.1}$  &    18.3 $\pm$     1.6 $\pm$     0.9 $_{-    1.6}^{+    0.8}$  &    12.2 $\pm$     1.9 $\pm$     0.6 $_{-    1.8}^{+    1.0}$ \\\hline
   5--6  &    39.1 $\pm$     3.4 $\pm$     1.9 $_{-   10.3}^{+    5.4}$  &    44.8 $\pm$     2.6 $\pm$     2.1 $_{-    7.6}^{+    3.9}$  &    32.8 $\pm$     2.1 $\pm$     1.5 $_{-    2.8}^{+    1.5}$  &    18.1 $\pm$     1.6 $\pm$     0.8 $_{-    1.2}^{+    0.6}$  &     7.8 $\pm$     1.5 $\pm$     0.4 $_{-    0.9}^{+    0.4}$ \\\hline
   6--7  &    28.8 $\pm$     2.9 $\pm$     1.4 $_{-    8.3}^{+    4.1}$  &    25.1 $\pm$     1.9 $\pm$     1.2 $_{-    3.9}^{+    2.0}$  &    22.3 $\pm$     1.7 $\pm$     1.0 $_{-    1.4}^{+    0.7}$  &    11.6 $\pm$     1.3 $\pm$     0.5 $_{-    0.5}^{+    0.3}$  &     5.2 $\pm$     1.2 $\pm$     0.2 $_{-    0.5}^{+    0.2}$ \\\hline
   7--8  &    21.9 $\pm$     2.4 $\pm$     1.0 $_{-    5.4}^{+    2.7}$  &    23.4 $\pm$     1.9 $\pm$     1.1 $_{-    3.5}^{+    1.8}$  &    16.3 $\pm$     1.4 $\pm$     0.8 $_{-    0.9}^{+    0.4}$  &     5.8 $\pm$     0.9 $\pm$     0.3 $_{-    0.1}^{+    0.1}$  &     5.4 $\pm$     1.2 $\pm$     0.3 $_{-    0.4}^{+    0.2}$ \\\hline
   8--9  &    22.9 $\pm$     2.4 $\pm$     1.1 $_{-    4.8}^{+    2.6}$  &    17.1 $\pm$     1.5 $\pm$     0.8 $_{-    2.0}^{+    1.0}$  &    12.4 $\pm$     1.2 $\pm$     0.6 $_{-    0.6}^{+    0.3}$  &     7.6 $\pm$     1.0 $\pm$     0.4 $_{-    0.2}^{+    0.1}$  &     4.3 $\pm$     1.0 $\pm$     0.2 $_{-    0.3}^{+    0.1}$ \\\hline
   9--10  &    12.8 $\pm$     1.8 $\pm$     0.6 $_{-    2.9}^{+    1.5}$  &    12.9 $\pm$     1.3 $\pm$     0.6 $_{-    1.2}^{+    0.6}$  &     9.8 $\pm$     1.1 $\pm$     0.5 $_{-    0.5}^{+    0.2}$  &     7.0 $\pm$     1.0 $\pm$     0.3 $_{-    0.2}^{+    0.1}$  &     1.2 $\pm$     0.5 $\pm$     0.1 $_{-    0.1}^{+    0.0}$ \\\hline
  10--11  &    10.3 $\pm$     1.6 $\pm$     0.5 $_{-    2.1}^{+    1.1}$  &     9.5 $\pm$     1.1 $\pm$     0.4 $_{-    0.9}^{+    0.5}$  &     4.3 $\pm$     0.7 $\pm$     0.2 $_{-    0.2}^{+    0.1}$  &     6.4 $\pm$     0.9 $\pm$     0.3 $_{-    0.2}^{+    0.1}$  &     2.6 $\pm$     0.8 $\pm$     0.1 $_{-    0.2}^{+    0.1}$ \\\hline
  11--12  &     8.6 $\pm$     1.5 $\pm$     0.4 $_{-    2.4}^{+    1.2}$  &    10.0 $\pm$     1.1 $\pm$     0.5 $_{-    0.9}^{+    0.5}$  &     4.4 $\pm$     0.7 $\pm$     0.2 $_{-    0.1}^{+    0.0}$  &     1.2 $\pm$     0.4 $\pm$     0.1 $_{-    0.0}^{+    0.0}$  & - \\\hline
  12--13  &     5.8 $\pm$     1.2 $\pm$     0.3 $_{-    0.9}^{+    0.5}$  &     5.8 $\pm$     0.9 $\pm$     0.3 $_{-    0.5}^{+    0.3}$  &     4.1 $\pm$     0.7 $\pm$     0.2 $_{-    0.1}^{+    0.0}$  & -  & - \\\hline
  13--14  &     4.4 $\pm$     1.0 $\pm$     0.2 $_{-    0.7}^{+    0.4}$  &     1.7 $\pm$     0.5 $\pm$     0.1 $_{-    0.1}^{+    0.1}$  &     2.6 $\pm$     0.5 $\pm$     0.1 $_{-    0.1}^{+    0.0}$  & -  & - \\\hline
  14--15  &     1.9 $\pm$     0.6 $\pm$     0.1 $_{-    0.3}^{+    0.2}$  &     4.9 $\pm$     0.8 $\pm$     0.2 $_{-    0.5}^{+    0.3}$  &     3.9 $\pm$     0.7 $\pm$     0.2 $_{-    0.3}^{+    0.1}$  & -  & - \\\hline
\end{tabular}
}
\end{center} 
\end{table}

 \begin{table}[!ht] 
\begin{center} 
\caption{\small{Double differential cross-section  $\threesmm$ 
    as a function of rapidity  and transverse momentum, in pb/(\gevc).
    The first uncertainty is statistical,
    the second is systematic, and the third is due to the unknown polarisation of the $\threes$.
    Regions where the number of events was not sufficient to perform a measurement
    are indicated with a dash.}
}\label{tab::xsec3s_corr} 
\resizebox{\textwidth}{!}{ 
\begin{tabular}{|c|c|c|c|c|c|} 
\hline 
 $\pt$ & $2.0<y<2.5$ & $2.5<y<3.0$ & $3.0<y<3.5$ & $3.5<y<4.0$ & $4.0<y<4.5$  \\ 
 \small{(\gevc)} & & &  &  &  \\ \hline
   0--1  &     7.0 $\pm$     1.5 $\pm$     0.3 $_{-    2.6}^{+    1.3}$  &     6.3 $\pm$     1.0 $\pm$     0.3 $_{-    1.0}^{+    0.6}$  &     3.1 $\pm$     0.6 $\pm$     0.1 $_{-    0.4}^{+    0.2}$  &     5.0 $\pm$     0.9 $\pm$     0.2 $_{-    0.9}^{+    0.5}$  & - \\\hline
   1--2  &    14.1 $\pm$     2.2 $\pm$     0.7 $_{-    5.3}^{+    2.6}$  &     5.6 $\pm$     0.9 $\pm$     0.3 $_{-    1.1}^{+    0.6}$  &    11.6 $\pm$     1.2 $\pm$     0.6 $_{-    1.3}^{+    0.7}$  &    12.7 $\pm$     1.4 $\pm$     0.6 $_{-    2.1}^{+    1.2}$  &    10.2 $\pm$     1.9 $\pm$     0.5 $_{-    2.6}^{+    1.4}$ \\\hline
   2--3  &    17.6 $\pm$     2.3 $\pm$     0.9 $_{-    5.3}^{+    2.7}$  &    22.3 $\pm$     1.8 $\pm$     1.1 $_{-    4.1}^{+    2.1}$  &    15.2 $\pm$     1.4 $\pm$     0.7 $_{-    1.6}^{+    0.8}$  &     6.7 $\pm$     1.0 $\pm$     0.3 $_{-    0.9}^{+    0.5}$  &     9.9 $\pm$     1.7 $\pm$     0.5 $_{-    2.1}^{+    1.2}$ \\\hline
   3--4  &    24.9 $\pm$     2.7 $\pm$     1.2 $_{-    7.7}^{+    4.0}$  &    17.6 $\pm$     1.6 $\pm$     0.8 $_{-    3.1}^{+    1.6}$  &    13.5 $\pm$     1.3 $\pm$     0.6 $_{-    1.6}^{+    0.8}$  &     6.8 $\pm$     1.0 $\pm$     0.3 $_{-    0.8}^{+    0.4}$  &     7.5 $\pm$     1.5 $\pm$     0.4 $_{-    1.3}^{+    0.7}$ \\\hline
   4--5  &    16.7 $\pm$     2.2 $\pm$     0.8 $_{-    5.1}^{+    2.6}$  &    17.5 $\pm$     1.6 $\pm$     0.8 $_{-    3.0}^{+    1.6}$  &     6.9 $\pm$     0.9 $\pm$     0.3 $_{-    0.6}^{+    0.3}$  &     6.1 $\pm$     0.9 $\pm$     0.3 $_{-    0.5}^{+    0.3}$  &     7.6 $\pm$     1.5 $\pm$     0.4 $_{-    1.2}^{+    0.6}$ \\\hline
   5--6  &    16.6 $\pm$     2.1 $\pm$     0.8 $_{-    4.6}^{+    2.4}$  &    21.3 $\pm$     1.8 $\pm$     1.0 $_{-    3.5}^{+    1.8}$  &    12.1 $\pm$     1.2 $\pm$     0.6 $_{-    1.1}^{+    0.6}$  &     7.8 $\pm$     1.1 $\pm$     0.4 $_{-    0.5}^{+    0.3}$  &     7.6 $\pm$     1.4 $\pm$     0.4 $_{-    0.9}^{+    0.5}$ \\\hline
   6--7  &    22.2 $\pm$     2.5 $\pm$     1.1 $_{-    5.6}^{+    3.0}$  &    19.1 $\pm$     1.7 $\pm$     0.9 $_{-    3.0}^{+    1.5}$  &     8.4 $\pm$     1.0 $\pm$     0.4 $_{-    0.6}^{+    0.3}$  &     7.1 $\pm$     1.0 $\pm$     0.3 $_{-    0.3}^{+    0.2}$  &     3.1 $\pm$     0.9 $\pm$     0.2 $_{-    0.3}^{+    0.1}$ \\\hline
   7--8  &    20.6 $\pm$     2.4 $\pm$     1.0 $_{-    5.4}^{+    2.7}$  &    10.5 $\pm$     1.2 $\pm$     0.5 $_{-    1.6}^{+    0.8}$  &     9.2 $\pm$     1.1 $\pm$     0.4 $_{-    0.6}^{+    0.3}$  &     5.2 $\pm$     0.9 $\pm$     0.3 $_{-    0.1}^{+    0.1}$  &     1.4 $\pm$     0.6 $\pm$     0.1 $_{-    0.1}^{+    0.1}$ \\\hline
   8--9  &    13.7 $\pm$     1.9 $\pm$     0.7 $_{-    3.3}^{+    1.7}$  &    10.7 $\pm$     1.2 $\pm$     0.5 $_{-    1.6}^{+    0.8}$  &     6.8 $\pm$     0.9 $\pm$     0.3 $_{-    0.3}^{+    0.1}$  &     2.4 $\pm$     0.6 $\pm$     0.1 $_{-    0.1}^{+    0.0}$  &     0.6 $\pm$     0.4 $\pm$     0.0 $_{-    0.0}^{+    0.0}$ \\\hline
   9--10  &    11.3 $\pm$     1.7 $\pm$     0.5 $_{-    2.5}^{+    1.3}$  &     6.9 $\pm$     1.0 $\pm$     0.3 $_{-    0.8}^{+    0.4}$  &     5.7 $\pm$     0.8 $\pm$     0.3 $_{-    0.3}^{+    0.2}$  &     2.5 $\pm$     0.6 $\pm$     0.1 $_{-    0.1}^{+    0.0}$  &     3.2 $\pm$     0.9 $\pm$     0.2 $_{-    0.1}^{+    0.1}$ \\\hline
  10--11  &     8.4 $\pm$     1.5 $\pm$     0.4 $_{-    2.0}^{+    1.0}$  &     5.5 $\pm$     0.9 $\pm$     0.3 $_{-    0.6}^{+    0.3}$  &     4.3 $\pm$     0.7 $\pm$     0.2 $_{-    0.2}^{+    0.1}$  &     2.6 $\pm$     0.6 $\pm$     0.1 $_{-    0.1}^{+    0.1}$  & - \\\hline
  11--12  &     8.7 $\pm$     1.4 $\pm$     0.4 $_{-    1.7}^{+    0.9}$  &     4.4 $\pm$     0.7 $\pm$     0.2 $_{-    0.3}^{+    0.2}$  &     3.2 $\pm$     0.6 $\pm$     0.2 $_{-    0.2}^{+    0.1}$  &     1.8 $\pm$     0.5 $\pm$     0.1 $_{-    0.1}^{+    0.0}$  & - \\\hline
  12--13  &     4.5 $\pm$     1.0 $\pm$     0.2 $_{-    0.9}^{+    0.4}$  &     3.2 $\pm$     0.6 $\pm$     0.2 $_{-    0.3}^{+    0.1}$  &     3.5 $\pm$     0.7 $\pm$     0.2 $_{-    0.1}^{+    0.1}$  & -  & - \\\hline
  13--14  &     2.4 $\pm$     0.7 $\pm$     0.1 $_{-    0.4}^{+    0.2}$  &     0.7 $\pm$     0.3 $\pm$     0.0 $_{-    0.1}^{+    0.0}$  &     2.1 $\pm$     0.5 $\pm$     0.1 $_{-    0.1}^{+    0.0}$  & -  & - \\\hline
  14--15  &     0.7 $\pm$     0.4 $\pm$     0.0 $_{-    0.1}^{+    0.1}$  &     1.5 $\pm$     0.4 $\pm$     0.1 $_{-    0.1}^{+    0.1}$  &     0.9 $\pm$     0.3 $\pm$     0.0 $_{-    0.0}^{+    0.0}$  & -  & - \\\hline
\end{tabular}
}
\end{center} 
\end{table}

\renewcommand{\arraystretch}{1.0}
\begin{table}[!ht] 
\begin{center} 
\caption{\small{
   Ratios of cross-sections
  $\twosmm$ and $\threesmm$  with respect to \nobreak{$\onesmm$}
 as a function of $\pt$ in the
range $2.0<y<4.5$, assuming no polarisation. The first uncertainty is statistical, 
the second is systematic and the third is due to the unknown polarisation
of the three states. 
}}\label{tab::ratios} 
\vspace{0.3cm} 
\resizebox{0.7\textwidth}{!}{ 
\begin{tabular}{c|c|c} 
\noalign{\smallskip}
 $\pt$ & $R^{2S/1S}$ & $R^{3S/1S}$ \\ 
 \small{(\gevc)} &  &  \\ 
\noalign{\smallskip}
\hline
\noalign{\smallskip}
   0--1  &   0.202 $\pm$   0.015 $\pm$   0.006 $\pm$   0.052 &   0.099 $\pm$   0.010 $\pm$   0.003 $\pm$   0.025\\ 
   1--2  &   0.192 $\pm$   0.009 $\pm$   0.005 $\pm$   0.051 &   0.089 $\pm$   0.006 $\pm$   0.003 $\pm$   0.024\\ 
   2--3  &   0.207 $\pm$   0.008 $\pm$   0.006 $\pm$   0.052 &   0.098 $\pm$   0.005 $\pm$   0.003 $\pm$   0.025\\ 
   3--4  &   0.247 $\pm$   0.010 $\pm$   0.007 $\pm$   0.056 &   0.099 $\pm$   0.006 $\pm$   0.003 $\pm$   0.023\\ 
   4--5  &   0.234 $\pm$   0.010 $\pm$   0.007 $\pm$   0.047 &   0.087 $\pm$   0.005 $\pm$   0.003 $\pm$   0.017\\ 
   5--6  &   0.305 $\pm$   0.013 $\pm$   0.009 $\pm$   0.058 &   0.136 $\pm$   0.007 $\pm$   0.005 $\pm$   0.023\\ 
   6--7  &   0.260 $\pm$   0.013 $\pm$   0.007 $\pm$   0.048 &   0.160 $\pm$   0.009 $\pm$   0.006 $\pm$   0.027\\ 
   7--8  &   0.268 $\pm$   0.015 $\pm$   0.008 $\pm$   0.048 &   0.162 $\pm$   0.011 $\pm$   0.006 $\pm$   0.027\\ 
   8--9  &   0.309 $\pm$   0.019 $\pm$   0.009 $\pm$   0.046 &   0.166 $\pm$   0.013 $\pm$   0.006 $\pm$   0.028\\ 
   9--10  &   0.303 $\pm$   0.022 $\pm$   0.009 $\pm$   0.045 &   0.187 $\pm$   0.016 $\pm$   0.007 $\pm$   0.032\\ 
\noalign{\smallskip}
\end{tabular}
}
\end{center} 
\end{table}

\clearpage

\bibliographystyle{LHCb}
\bibliography{main}

\end{document}